\documentclass[a4paper,fleqn,usenatbib]{mnras}
\usepackage{newtxtext,newtxmath}
\usepackage[T1]{fontenc}
\usepackage{ae,aecompl}
\usepackage{color}
\usepackage{enumerate}
      
\usepackage{graphicx}	
\usepackage{amsmath}	
\usepackage{amssymb}	

\DeclareMathOperator{\sech}{sech}   
\def\msun{\,M_\odot}
\def\pc{\,{\rm pc}}
\def\kpc{\,{\rm kpc}}
\def\kms{\,{\rm km\,s^{-1}}}
\def\myr{\,{\rm Myr}}
\def\gyr{\,{\rm Gyr}}
\def\feh{\left[{\rm Fe/H}\right]}
\def\e{{\rm e}}
\def\rd{{\rm d}}

\def\afe{\left[{\rm \alpha/Fe}\right]}
\def\P5{P1$\sigma$}
\def\R2{R2$\sigma$}
\def\U1o{U1$\sigma\zeta$*}
\def\WA1{X5$\zeta$*}
\def\PA1{T5$\zeta$*}
\def\C3b{C$\beta$2s5}
\def\M2{M$\beta$1s5}
\def\V6{V$\alpha$9s8$\lambda\zeta$*}

\title[Structural evolution of thin and thick discs]
{The structural evolution of galaxies with both thin and thick discs}
\author[M. Aumer \& J. Binney]
{Michael Aumer \thanks{E-mail:Michael.Aumer@physics.ox.ac.uk (MA)}
and James Binney\\
Rudolf Peierls Centre for Theoretical Physics, 1 Keble Road, Oxford, OX1 3NP, UK}

\date{Accepted 2017 May 22. Received 2017 May 22; in original form 2017 March 24}
\pubyear{2017}
\begin{document}
\label{firstpage}
\pagerange{\pageref{firstpage}--\pageref{lastpage}} 
\maketitle

\begin{abstract}

We perform controlled $N$-body simulations of disc galaxies growing within
live dark matter (DM) haloes to present-day galaxies that contain both thin
and thick discs. We consider two types of models: a) thick-disc initial
conditions to which stars on near-circular orbits are continuously added
over $\sim10\gyr$, and b) models in which the birth velocity dispersion of
stars decreases continuously over the same time-scale. We show that both
schemes produce double-exponential vertical profiles similar to that of the
Milky Way (MW). We indicate how the spatial age structure of galaxies
can be used to discriminate between scenarios. We show that the presence of
a thick disc significantly alters and delays bar formation and thus makes possible
models with a realistic bar {\it and} a high baryon-to-DM mass ratio in the
central regions, as required by microlensing constraints. We examine how the
radial mass distribution in stars and DM is affected by disc growth and
non-axisymmetries. We discuss how bar buckling shapes the vertical age
distribution of thin- and thick-disc stars in the bar region. The extent
to which the combination of observationally motivated inside-out growth
histories and cosmologically motivated dark halo properties leads to the
spontaneous formation of non-axisymmetries that steer the models towards
present-day MW-like galaxies is noteworthy.

\end{abstract}


\section{Introduction}

The vertical density profile of stars in the Milky Way (MW) is fitted well by a sum
of two exponentials \citep{gilmore}. \citet{juric} find scaleheights of $\sim900\pc$ for
the {\it geometrical thick disc} and $\sim300\pc$ for the {\it thin disc}. Similarly, the vertical 
surface brightness profiles of the majority of bright edge-on spiral galaxies show thin and thick 
components \citep{yoachim}. Studies of the chemical abundances of solar neighbourhood (Snhd) stars 
reveal that populations with hotter vertical kinematics and thus larger scaleheights have abundances 
of $\alpha$ elements relative to iron ($\afe$) that are larger than those of populations with small 
scaleheights and comparable iron abundances $\feh$ \citep{fuhrmann, bensby}. 

In the $\afe-\feh$ plane, the low- and high-$\alpha$ populations are generally found to separate
into fairly distinct sequences. Higher $\afe$ indicates shorter chemical enrichment time-scales
and age determinations find systematically older ages for stars of the $\alpha$ enhanced  {\it 
chemical thick disc} \citep{masseron, martig}: they are generally found to be older
than 8 Gyr. The distinct sequences have motivated models in which the two components 
formed in two temporally separated phases in very different conditions (e.g. \citealp{chiappini}),
but can also be explained as the result of continuous star formation and chemical enrichment 
\citep{sb09b}.

It has now become clear that chemical and geometrical definitions of the thick disc yield different
results. Whereas \citet{juric}, who determined the density of all stars in the Snhd independent of chemistry,
found that the geometrically thick disc has a longer radial scalelength than the thin disc,
high-$\afe$ stars are found to form a thicker, but more centrally concentrated component than low-$\afe$ stars 
\citep{bovy, hayden}. A scenario in which the disc forms inside-out and each mono-age population
{\it flares}, i.e. is thicker at outer than at inner radii, could potentially explain these observations
\citep{minchev, ralph2017}. Flaring can be caused by radial migration of stars \citep{sb12, roskar}
or by vertical heating of the outer disc through satellite interactions \citep{kazantzidis} or misaligned 
gas infall \citep{jiang}.

\citet{abs16a, abs16b} (hereafter Papers 1 and 2) presented $\sim100$
idealized $N$-body models of disc galaxies growing within live dark matter
(DM) haloes over $\sim 10\gyr$. These models covered a large variety of star
formation and radial growth histories and most of them followed the
assumption that all stars are born on near-circular orbits as in the MW
today. Structural and kinematical properties of the MW's thin disc,
such as an exponential profile with scaleheight $\sim300\pc$ or the local
age-velocity dispersion relations could be reproduced if giant molecular
clouds (GMCs) were included. Additionally, bars of
comparable size and structure to that of the MW formed in these models.
However, none of the models produced a realistic thick disc. The
conclusion was that additional sources of heating were required early in the
disc's life, prior to the onset of thin-disc  formation.

In this paper, we create models similar to those of Paper 1 that, in addition to a realistic 
thin disc, also contain an appropriate $\sim 10 \gyr$ old thick disc.  Demanding the presence
of an old thick component makes it much harder to steer a model to a configuration consistent 
with current data, in part because the number of observations that need to be explained 
simultaneously increases roughly twofold. Moreover, the thick disc must be steered into its 
present form by adjusting the conditions at the onset of disc formation, which will modify the
subsequent formation of the thin-disc component by altering spiral and bar structures and their
interaction with the dark halo. These changes to the thin disc and dark halo will themselves
modify the appearance of the current thick disc. Moreover, the old chemical thick disc of the  
MW is centrally concentrated and should be important in the central $\sim5\kpc$, a region 
dominated by the Galactic bar \citep{portail}, a structure that is supposed to have formed 
from a rather cold disc.

To create a thick disc we follow two approaches. In one scheme, we assume that the velocity
dispersions of newborn stars decline with time. This is motivated by observations
of redshift $z_{\rm rs}\sim2$ galaxies, which show a high fraction of galaxies with H$\alpha$ kinematics 
consistent with ordered rotation, but significantly higher velocity dispersions than today's disc 
galaxies \citep{sins} and of galaxies at lower redshifts that indicate a continuous decline of gas 
velocity dispersion with decreasing redshift (\citealp{kassin, wisn}, but see \citealp{diteodoro} for
a different conclusion). Such a declining birth dispersion
has also been found in hydrodynamical cosmological simulations of disc galaxy formation \citep{bird, grand},
and can be understood in terms of galaxies that gradually become less gas-rich and are characterized 
by gravitationally driven turbulence that decays \citep{forbes}.

In an alternative scheme, we model thin+thick disc systems by creating thick
initial conditions and growing thin discs within them. These models are thus
representations of two-phase formation scenarios. We here do not model the
formation of the thick disc prior to redshift $z_{\rm rs}\sim2$, but several
overlapping scenarios envisage the production of such an object: heating of
an initially thinner disc by a merger (e.g. \citealp{quinn}); formation in
early gas-rich mergers \citep{brook}; the formation in a clumpy, turbulent
disc \citep{bournaud}.  It should be noted that the latter two scenarios
could also be fitted into the picture of declining birth dispersions that
constitutes our alternative modelling scheme.

In this paper, we concentrate on the setup of our models and on their
structural evolution. We explore how models with double-exponential
vertical profiles, circular speed curves like that of the MW and bars of
appropriate structure can be constructed.  We examine how the presence of a
thick disc changes the preferred density of the dark halo and influences the
evolution of the thin disc and the formation of a bar. We show that different
scenarios for the formation of the thick disc leave signatures in the
current distribution of age with the $(R,z)$ plane. A companion paper
(\citealp{abs17}, hereafter Paper 4) focuses on disc heating and radial migration
in the models.

Our paper is structured as follows. 
Section \ref{sec:simulations} describes the setup and parameters of our simulations.
Section \ref{sec:vert} discusses the evolution of vertical density profiles and how 
this shapes the final age structure of the disc. Section \ref{sec:rad} analyses 
the radial distribution of dark and baryonic mass components and Section \ref{sec:bar}
illustrates the evolution and structure of bars in the presence of thick discs.
Section \ref{sec:discuss} discusses the successes and problems of our models.
Section \ref{sec:conclude} concludes.

\section{Simulations}
\label{sec:simulations}

The simulations analysed in this paper are similar to the models presented
in Paper 1. These are simulations of growing disc galaxies within non-growing
live DM haloes. They are run with the Tree code GADGET-3, last described in \citet{gadget2}.
We focus on standard-resolution models, which contain $N=5\times10^6$ particles in the 
DM halo and a similar number of particles in the final stellar system.
We here rely on collisionless simulations; models that contain an isothermal
gas component were discussed in Papers 1 and 2 and shown to differ only mildly
from collisionless models in terms of the structure and kinematics of the 
stellar component. In addition, and crucially, all simulations contain a population of 
short-lived, massive particles representing GMCs. Papers 1 and 2 demonstrated
the importance of GMC heating in creating thin-disc components with realistic 
vertical structure as non-axisymmetric structure contributes little to vertical
disc heating (see also \citealp{sellwood13}).

\begin{table*}
  \caption{An Overview over the different ICs and their parameters:
           {\it 1st Column}:  IC name;
           {\it 2nd Column}:  IC disc mass $M_{\rm disc,i}$;
           {\it 3rd Column}:  IC bulge mass $M_{\rm bulge,i}$;
           {\it 4th Column}:  number of baryonic particles $N_{\rm b,i}$ in the ICs;
           {\it 5th Column}:  concentration parameter for IC DM halo, $c_{\rm halo}$;
           {\it 6th Column}:  IC DM halo scalelength $a_{\rm halo}$;
           {\it 7th Column}:  IC radial disc scalelength $h_{R, {\rm disc}}$;
           {\it 8th Column}:  IC vertical disc scaleheight $z_{0, {\rm disc}}$;
           {\it 9th Column}:  IC bulge scalelength $a_{\rm bulge}$; 
           {\it 10th Column}: IC bulge flattening $s$; 
           {\it 11th Column}: IC disc velocity ellipsoid ratio $\sigma_R^2/\sigma_z^2$; 
           {\it 12th Column}: IC bulge rotation. 
}
  \begin{tabular}{@{}cccccccccccc@{}}\hline
    1st    &  2nd           & 3rd            & 4th         & 5th         & 6th         & 7th             & 8th             & 9th          & 10th   & 11th & 12th\\
    {Name} &{$M_{\rm disc,i}$}&{$M_{\rm bulge,i}$} &{$N_{\rm b,i}$}&{$c_{\rm halo}$}&{$a_{\rm halo}$}&{$h_{R,{\rm disc}}$}&{$z_{0,{\rm disc}}$}&{$a_{\rm bulge}$}& {$s$}& $\sigma_R^2/\sigma_z^2$ & rotation\\ 
           &{$[10^9\msun]$} & {$[10^9\msun]$} &             &              &{$[\kpc]$}    & {$[\kpc]$}    & {$[\kpc]$}       & {$[\kpc]$}    &      & disc  &  bulge   \\ \hline

    Y      &$5$             &  --             &$500\,000$   &    9         &  30.2       &  1.5           &   0.10           &   --         &  --  &  2.0 &   --\\\hline

    P      &$15$            &  --             &$1\,500\,000$&    9         &  30.2       &  2.5           &   1.75           &   --         & --   &  1.0 &   --\\
    Q      &$25$            &  --             &$2\,500\,000$&    6.5       &  37.9       &  2.5           &   1.75           &   --         & --   &  1.0 &   --\\
    R      &$15$            &  --             &$1\,500\,000$&    6.5       &  37.9       &  2.5           &   1.75           &   --         & --   &  1.0 &   --\\
    U      &$25$            &  --             &$2\,500\,000$&    7.5       &  34.4       &  2.0           &   1.70           &   --         & --   &  1.0 &   --\\
    T      &$20$            &  --             &$2\,000\,000$&    9         &  30.2       &  2.0           &   1.70           &   --         & --   &  1.8 &   --\\\hline
  
    C      & --             &  $5$            &$500\,000$   &    9         &  30.2       &  --           &   --            &   0.45        & 1.15 &  --  &   no\\
    K      & --             &  $15$           &$1\,500\,000$&    9         &  30.2       &  --           &   --            &   1.51        & 2.0  &  --  &   yes\\
    M      & --             &  $5$            &$500\,000$   &    9         &  30.2       &  --           &   --            &   1.51        & 2.0  &  --  &   yes\\
    O      & --             &  $15$           &$1\,500\,000$&    9         &  30.2       &  --           &   --            &   1.51        & 3.0  &  --  &   yes\\
    V      & --             &  $5$            &$500\,000$   &    6.5       &  37.9       &  --           &   --            &   1.51        & 2.0  &  --  &   yes\\\hline

    W      &$20$            &  $5$            &$2\,500\,000$&    7.5       &  34.4       &  2.5           &   1.70           &   0.46        & 1.15 &  1.0 &   no\\
    X      &$20$            &  $5$            &$2\,500\,000$&    7.5       &  34.4       &  2.0           &   1.70           &   0.70        & 1.15 &  1.8 &   no\\\hline
    
\hline
  \end{tabular}
  \label{ictable}
\end{table*}

\subsection{Initial conditions}

Table~\ref{ictable} gives an overview of the initial conditions (ICs) of our models.
The ICs were created using the GALIC code \citep{yurin}. 
We refer the reader to Paper 1 for a full description of the IC creation process and
here focus on how the ICs used in this paper differ from each other. All models 
discussed start with a spherical DM halo with a \citet{hernquist} profile
\begin{equation}
\rho_{\rm{DM}}(r)={{M_{\rm{DM}}}\over{2\pi}} {{a}\over{r\left(r+a\right)^3}}.
\end{equation}
The total mass of all ICs is $M_{\rm tot}=10^{12}\msun$, of which $M_{\rm b,i}=M_{\rm disc,i}+M_{\rm bulge,i}$
is in a stellar component and the rest $M_{\rm{DM}}$ is in DM.
The inner density profile is adjusted so that it is similar to an NFW profile with
concentration $c_{\rm halo}=6-9$ and thus scale radii are in the range $a=30-40\kpc$.
The kinematics of the DM particles are initially isotropic
(i.e. have equal velocity dispersions $\sigma_{r}=\sigma_{\phi}=\sigma_{\theta}$).
Each DM halo studied here is resolved with $N_{\rm DM}=5\,000\,000$ particles.

ICs contain either a disc or a bulge component or both. IC discs have a mass in the range 
$M_{\rm{disc,i}}=5-25\times 10^9 \msun$ and a density profile of the form
\begin{equation}
{\rho_{\rm{disc,i}}(R,z)} = {{M_{\rm{disc,i}}}\over{4\pi {z_{0,{\rm
disc}}}{h_{R,{\rm disc}}}^2}} {\sech^2 \left({z}\over{z_{0,{\rm
disc}}}\right)} {\exp\left(-{R}\over{h_{R,{\rm disc}}}\right)},
\end{equation}
where $h_{R,{\rm disc}}=1.5-2.5\kpc$ is the exponential disc scalelength. Radially constant
isothermal vertical profiles with scaleheights $z_{0, {\rm disc}}=0.1-1.7\kpc$ are assumed.
The ratios of the radial to vertical velocity dispersions are also radially constant in the IC discs
and have values in the range ${\sigma_R^2}/{\sigma_z^2}=1.0-2.0$.

IC bulge components are set up with Hernquist density profiles with scalelengths $a_{\rm bulge}=0.4-1.5\kpc$
that are distorted to be oblate spheroids with axis ratios in the range $s=1-3$ as
\begin{equation}
\rho_{\rm bulge}(R,z)=s \rho_{\rm Hernquist}\left(\sqrt{R^2+s^2z^2}\right).
\end{equation}
They are supposed to model either compact, non-rotating bulges or rotating spheroidal components.
Following section 3.2 of \citet{yurin}, the rotation of the axisymmetric bulge components is 
controlled via the \citet{satoh} parametrization
\begin{equation}
\left<v_{\phi}\right>^2 = k^2 \left( \left< v_{\phi}^2\right> -\sigma_R^2\right).
\end{equation}
Non-rotating bulges assume $k=0$, whereas rotating bulges are modelled as isotropic rotators with
$k=1$.

Stellar particles in the ICs have a particle mass $m=1\times10^4 \msun$ and 
our ICs thus contain $N_{\rm b,i}=5-25\times 10^5$ stellar particles.
As in Paper 1, the applied force softening lengths for the given resolutions 
are $\epsilon_{\rm b}=30\pc$ for baryonic particles (including GMCs) and 
$\epsilon_{\rm DM}=134\pc$ for DM particles.

The first letter of a model's name specifies its ICs according
to the scheme laid out in Table \ref{ictable}. In summary:
\begin{enumerate}[(i)]
\item{Y ICs have a compact and thin baryonic disc;}
\item{P, Q, R, U and T ICs  contain more extended and thicker discs, with varying size, 
thickness, mass and DM halo concentration;}
\item{C ICs have a compact non-rotating bulge;}
\item{W and X ICs combine a compact non-rotating bulge with a thick disc;}
\item{K, M, O and V ICs contain rotating oblate spheroids that crudely represent elliptical galaxies.}
\end{enumerate}

\subsection{Growing the disc}

To simulate the growth of galaxies over cosmological time-scales, stellar particles
with a particle mass $m=1\times10^4 \msun$
are continuously added to the simulations. As we are interested in the coevolution of thin
and thick discs, we pursue two different ideas, which were already touched upon in
Paper 1. a) We assume that a thick and rotating stellar component was formed early
on in a galaxy's history (e.g. through a merger), represent this component with a
thick disc or rotating spheroid in our ICs and over the timespan of the simulation
add star particles to the system on near-circular orbits. b) We assume that the 
birth velocity dispersions of stars have been continuously declining over the 
history of a galaxy and add stellar populations with continuously decreasing
dispersions.

\subsubsection{Evolution of input velocity dispersion}

In case a), the young stellar populations are assigned low birth velocity dispersions
in the range $\sigma_0=\sigma_1=6-10 \kms$ in all three directions $R$, $\phi$ and $z$ as observed 
in the MW (see blue stars in \citealp{ab09}). The mean rotation velocity $v_{\phi}(R)$ at
radius $R$ is set to the circular velocity $v_{\rm circ}=\sqrt{{a_R(R)} R}$, where $a_R(R)$
is the azimuthal average of the radial gravitational acceleration, $\partial\Phi/\partial R$.
As was shown in Papers 1 and 2, the in-plane dispersions quickly adjust to higher values 
and an appropriate ratio $\sigma_{\phi}/\sigma_{R}$ due to spiral and bar heating. 

In case b) we choose a value of $\sigma_0$ that is declining with time. Paper 1 had applied
\begin{equation}
\sigma_0(t)=\left(6+30\e^{-t/1.5\!\gyr}\right)\kms,
\end{equation}
which provided significantly too few hot stars. So, here we consider different functional forms:
\begin{equation}\label{atan}
\sigma_0(t)=\sigma_1 \left[ {\arctan \left( {t_1-t} \over {1\gyr}  \right)} + {{\pi} \over {2}} \right]+ 
\sigma_2 \,\,\, {\rm (type \, atan)}
\end{equation}
or
\begin{equation}\label{plaw}
\sigma_0(t)=\sigma_1 \left( {t+t_1} \over {t_2} \right)^{-\iota} -\sigma_2 \,\,\, {\rm (type \, plaw)}
\end{equation}

\begin{figure}
\vspace{-0.cm}
\includegraphics[width=8cm]{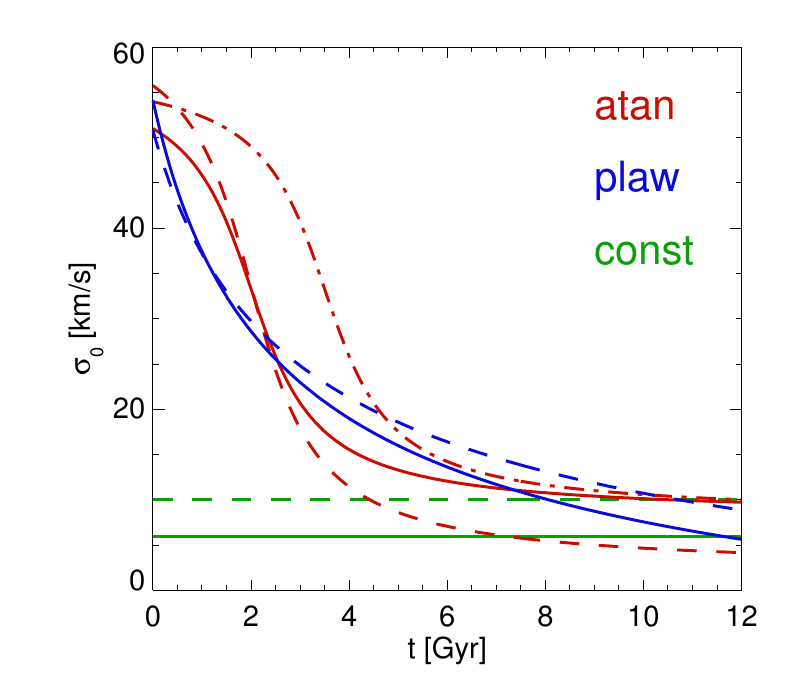}\\
\caption
{Evolution of the input velocity dispersion parameter $\sigma_0$ with simulation time $t$ for various scenarios.
Green lines are for type `const' (solid for $\sigma_0=6\kms$, dashed for $\sigma_0=10\kms$),
blue lines are for type `plaw' (solid for $\iota=0.56$, dashed for $\iota=0.47$, other parameters as in Table \ref{modeltable})
and red lines are for type `atan' (solid for $\sigma_1=16\kms$, $\sigma_2=8\kms$, $t_1=2\gyr$; dashed for 
$\sigma_1=20\kms$, $\sigma_2=2\kms$, $t_1=2\gyr$; dot-dashed for $\sigma_1=16\kms$, $\sigma_2=8\kms$, $t_1=3.5\gyr$). }
\label{sig0}
\end{figure}

In observations of high-redshift galaxies, it is common to study kinematics of the H$\alpha$
emission line and to assign a single characteristic velocity dispersion $\sigma_{\rm{H\alpha}}$ to each galaxy. Applying
such a procedure, \citet{wisn} find a dependence of observed H$\alpha$ dispersions 
$\sigma_{\rm{H\alpha}}\propto (1+z_{\rm rs})$, where $z_{\rm rs}$ is redshift (see also \citealp{kassin}).
The decline of our input velocity dispersion parameter $\sigma_0(t)$ for simulations of type
`plaw' (Equation \ref{plaw}) resulted from a rough approximation to this proportionality under the crude assumption
that the kinematics of young stars follow H$\alpha$ kinematics. Note that this trend is observed
for populations of galaxies at varying redshifts $z_{\rm rs}$. The spread in $\sigma_{\rm{H\alpha}}$
for each given redshift range in \citet{wisn} is significant and the expected formation histories
for galaxies are diverse, so that it is reasonable to assume that individual galaxies can show
evolutions of velocity dispersions very different from $\sigma_{\rm{H\alpha}}\propto (1+z_{\rm rs})$.
We thus also test decline histories of type `atan' (Equation \ref{atan}), which represents a scenario in which the transition 
from hot formation to cold formation is faster, yielding rather distinct hot and cold phases.
Figure \ref{sig0} visualizes the difference between assumed $\sigma_0$ histories.

We assume that $\sigma_0$ is radially constant and always set $\sigma_z(R,t)=\sigma_0(t)$. Resolution
effects prevent high-$z_{\rm rs}$ observations from constraining the radial dependence of $\sigma_{\rm{H\alpha}}$
reliably, but the following two observational findings motivate this simple modelling approach.
a) \citet{newman} find that, if one characteristic value is assigned to each galaxy, $\sigma_{\rm{H\alpha}}$ 
depends little on the size of the galaxy. b)\citet{jones} show radial $\sigma_{\rm{H\alpha}}(R)$
profiles of lensed galaxies and these show various shapes and high dispersions at outer radii.

In the case of high input dispersions, it is not desirable to use $\sigma_{\phi}/\sigma_{R}=1$
for an input, as this creates velocity distributions out of equilibrium. When $\sigma_0<10\kms$,
heating by non-axisymmetries is very efficient for young stars (see Paper 2) and the dispersions will
quickly adjust to an appropriate ratio $\sigma_{\phi}/\sigma_{R}$, but heating after birth will play 
a minor role for high $\sigma_0$. Since $\sigma_{\phi}/\sigma_{R}$ depends on the shape of the rotation curve
and the radial profile $\sigma_{R}(R)$
(see e.g. \citealp{gd2}, section 4.4.3), we apply a simple approach:
We assume $\sigma_{\phi}=\sqrt{0.5}\sigma_R$ and an asymmetric drift correction for $\left<v_{\phi}\right>$
derived from Equation 4.228 in \citet{gd2}. We find that $\sigma_{\phi}/\sigma_{R}$ for these assumptions
only adjusts mildly after insertion.

\tabcolsep=4.pt
\begin{table*}
\vspace{-0cm}
  \caption{List of models analysed in this paper. 
           {\it 1st Column}: model name;
           {\it 2nd Column}: initial conditions;
           {\it 3rd Column}: final total baryonic mass $M_{\rm f}$ 
           (including initial baryonic mass);
           {\it 4th Column}: final time $t_{\rm f}$;   
           {\it 5th Column}: initial disc scalelength $h_{R, {\rm i}}$;
           {\it 6th Column}: final disc scalelength $h_{R, {\rm f}}$;
           {\it 7th Column}: scalelength growth parameter $\xi$;
           {\it 8th Column}: type of SFH law;
           {\it 9th Column}: exponential decay time-scale $t_{\rm SFR}$ for the star formation rate;
           {\it 10th Column}: radial-to-vertical dispersion ratio for inserted particles $\lambda$;
           {\it 11th Column}: prescription for initial velocity dispersion for inserted stellar particles, {$\sigma_0(t)$};
           {\it 12th-16th Column}: parameters $\sigma_1$, $\sigma_2$, $t_1$, $t_2$ and $\iota$, which determine 
           {$\sigma_0(t)$};
           {\it 17th Column}: GMC star formation efficiency $\zeta$.}
  \begin{tabular}{@{}ccccccccccccccccc@{}}\hline
1st   & 2nd  &  3rd       & 4th        & 5th          & 6th           &7th    & 8th     & 9th        & 10th      & 11th       & 12th       & 13th      & 14th      & 15th    & 16th    & 17th    \\
{Name}&{ICs} & {$M_{\rm f}$}&{$t_{\rm f}$}&{$h_{R,{\rm i}}$}&{$h_{R,{\rm f}}$}&{$\xi$}&{SF type}&{$t_{\rm SFR}$}&{$\lambda$}&{$\sigma_0$}&{$\sigma_1$}&{$\sigma_2$}&{$t_1$}   &{$t_2$}   &{$\iota$}&{$\zeta$}\\ 
      & &{$[10^{10}\msun]$}&{$[\gyr]$}  &{$[\kpc]$}    &{$[\kpc]$}     &       &         &{$[\gyr]$}  &           & type       &{$[\kms]$}  &{$[\kms]$} &{$[\gyr]$}&{$[\gyr]$}&         &  \\ \hline
   
Y1    & Y    & 5          & 10         & 1.5          & 4.3           & 0.5   & 1       & 8.0        & 1.0       & const      & 6         & --         & --       & --       & --      &  0.08   \\ 
Y1$\zeta$- & Y & 5        & 10         & 1.5          & 4.3           & 0.5   & 1       & 8.0        & 1.0       & const      & 6         & --         & --       & --       & --      &  0.04   \\ 
Y2    & Y    & 5          & 10         & 2.5          & 2.5           & 0.0   & 1       & 8.0        & 1.0       & const      & 6         & --         & --       & --       & --      &  0.08   \\ 
\hline
P1$\zeta$-& P& 5          & 10         & 1.5          & 4.3           & 0.5   & 1       & 8.0        & 1.0       & const      & 6         & --         & --       & --       & --      &  0.04   \\
P1s6  & P    & 5          & 10         & 1.5          & 4.3           & 0.5   & 0       & --         & 1.0       & const      & 6         & --         & --       & --       & --      &  0.08   \\
P1$\sigma$& P& 5          & 10         & 1.5          & 4.3           & 0.5   & 1       & 8.0        & 1.0       & const      & 10        & --         & --       & --       & --      &  0.08   \\
P2    & P    & 5          & 10         & 2.5          & 2.5           & 0.0   & 1       & 8.0        & 1.0       & const      & 6         & --         & --       & --       & --      &  0.08   \\
Q1    & Q    & 6          & 10         & 1.5          & 4.3           & 0.5   & 1       & 8.0        & 1.0       & const      & 6         & --         & --       & --       & --      &  0.08   \\
Q1$\zeta$-& Q& 6          & 10         & 1.5          & 4.3           & 0.5   & 1       & 8.0        & 1.0       & const      & 6         & --         & --       & --       & --      &  0.04   \\
R2$\sigma$& R& 5          & 10         & 2.5          & 2.5           & 0.0   & 1       & 8.0        & 1.0       & const      & 10        & --         & --       & --       & --      &  0.08   \\
U1    & U    & 6          & 10         & 1.5          & 4.3           & 0.5   & 1       & 8.0        & 1.0       & const      & 6         & --         & --       & --       & --      &  0.08   \\
U1$\sigma\zeta$*& U& 6    & 10         & 1.5          & 4.3           & 0.5   & 1       & 8.0        & 1.0       & const      & 10        & --         & --       & --       & --      &  0.06   \\
W1    & W    & 6          & 10         & 1.5          & 4.3           & 0.5   & 1       & 8.0        & 1.0       & const      & 6         & --         & --       & --       & --      &  0.08   \\
X5$\zeta$*& X& 6          & 10         & 1.5          & 3.5           & 0.5   & 1       & 8.0        & 1.0       & const      & 6         & --         & --       & --       & --      &  0.06   \\
T5$\zeta$*& T& 5.5        & 10         & 1.5          & 3.5           & 0.5   & 1       & 8.0        & 1.0       & const      & 6         & --         & --       & --       & --      &  0.06   \\
\hline
K2    & K    & 5          & 10         & 2.5          & 2.5           & 0.0   & 1       & 8.0        & 1.0       & const      & 6         & --         & --       & --       &  --     &  0.08   \\
O2    & O    & 5          & 10         & 2.5          & 2.5           & 0.0   & 1       & 8.0        & 1.0       & const      & 6         & --         & --       & --       & --      &  0.08   \\
\hline\hline
C$\alpha$2& C& 5          & 10         & 2.5          & 2.5           & 0.0   & 1       & 8.0        & 1.0       & atan       & 20        & 2          & 2.0      & --       & --      &  0.08   \\
C$\beta$2s5& C& 5         & 12         & 2.5          & 2.5           & 0.0   & 1       & 8.0        & 1.0       & plaw       & 51        & 15         & 1.57     & 2.7      & 0.56    &  0.08   \\

M$\alpha$1& M& 5          & 10         & 1.5          & 4.3           & 0.5   & 1       & 8.0        & 1.0       & atan       & 16        & 8          & 2.0      & --       & --      &  0.08   \\
M$\alpha$1$\zeta$*& M& 5  & 10         & 1.5          & 4.3           & 0.5   & 1       & 8.0        & 1.0       & atan       & 16        & 8          & 2.0      & --       & --      &  0.06   \\
M$\beta$1s5& M& 5         & 12         & 1.5          & 4.3           & 0.5   & 1       & 8.0        & 1.0       & plaw       & 51        & 15         & 1.57     & 2.7      & 0.47    &  0.08   \\
M$\alpha$8s7& M& 5        & 12         & 1.0          & 4.3           & 0.6   & 2       & 12.0       & 1.0       & atan       & 16        & 8          & 3.5      & --       & --      &  0.08   \\

V$\alpha$1& V& 5          & 10         & 1.5          & 4.3           & 0.5   & 1       & 8.0        & 1.0       & atan       & 16        & 8          & 2.0      & --       & --      &  0.08   \\
V$\alpha$5$\lambda$& V& 5 & 10         & 1.5          & 3.5           & 0.5   & 1       & 8.0        & 1.25      & atan       & 16        & 8          & 2.0      & --       & --      &  0.08   \\
V$\alpha$8s5& V& 6        & 12         & 1.0          & 4.3           & 0.6   & 1       & 8.0        & 1.0       & atan       & 16        & 8          & 2.0      & --       & --      &  0.08   \\
V$\alpha$8s7& V& 6        & 12         & 1.0          & 4.3           & 0.6   & 2       & 12.0       & 1.0       & atan       & 16        & 8          & 3.5      & --       & --      &  0.08   \\
V$\beta$8s5& V& 6         & 12         & 1.0          & 4.3           & 0.6   & 1       & 8.0        & 1.0       & plaw       & 51        & 15         & 1.57     & 2.7      & 0.47    &  0.08   \\
V$\alpha$9s7$\lambda\zeta$*& V& 6 &12  & 1.0          & 3.5           & 0.6   & 2       & 12.0       & 1.25      & atan       & 16        & 8          & 3.5      & --       & --      &  0.06   \\
V$\alpha$9s8$\lambda\zeta$*& V& 6 &12  & 1.0          & 3.5           & 0.6   & 1       & 6.0        & 1.25      & atan       & 16        & 8          & 2.0      & --       & --      &  0.06   \\
\hline
  \end{tabular}
  \label{modeltable}
\end{table*}

The choice for the ratio $\sigma_z/\sigma_{R}$ is also unclear. For high-$z_{\rm rs}$ discs, \citet{genzel}
studied $\sigma_{\rm{H\alpha}}$ as a function of disc inclination. Edge-on galaxies would thus be dominated
by in-plane dispersions, whereas face-on galaxies would be dominated by vertical dispersions. They found
mild indications for $\sigma_R>\sigma_z$, but \citet{wisn} could not confirm this trend with a larger
sample of galaxies. From the perspective of the thick-disc stars in the MW, \citet{piffl} find 
$\sigma_z\approx\sigma_{R}$. We test values $\sigma_R=\lambda\sigma_0$ with $\lambda=1-1.3$.

As we place all newly added star particles in the midplane $z=0$, the measured vertical dispersion 
$\sigma_z$ of a young stellar component is smaller than $\sigma_0$, as the midplane is at the bottom
of the vertical potential well and particles quickly lose kinetic energy moving away from it.

\subsubsection{Star formation history}

The star formation rate (SFR) is either constant (type 0) or
\begin{equation}
{\rm SFR}(t)={\rm SFR}_0 \times \exp({-t/t_{\rm SFR}})\,\,\, {\rm {(type \,1)}},
\end{equation}
or 
\begin{equation}
{\rm SFR}(t)={\rm SFR}_0 \times \exp({-t/t_{\rm SFR}}-{0.5\gyr/t})\,\,\, {\rm {(type \,2)}},
\end{equation}
with $t_{\rm SFR}=6-12\gyr$. Our simulations run for a total time of $t_{\rm f}=10-12\gyr$ 
and the constant ${\rm SFR}_0$ is adjusted to produce at $t_{\rm f}$ a target final baryonic
mass $M_{\rm f}$ in the range $5-6\times10^{10}\msun$, including the mass of the stars in the ICs. 
Paper 1 showed that for the range of DM density profiles adopted for our models, galaxy masses 
$M_{\rm f}$ in this range provide the right level of  self-gravity to explain the age-velocity 
dispersion relation of the Snhd. \citet{mcmillan} favours similar Galaxy masses for his MW
mass models.

\subsubsection{Radial growth history}

Particles are added randomly distributed in azimuth every five Myr 
with an exponential radial density profile $\Sigma_{\rm SF}(R)\propto\exp(-R/h_R(t))$.
The scalelength $h_R(t)$ of the newly added particles grows in time as
\begin{equation}\label{eq:hRt}
h_R(t)=h_{R,\rm i}+(h_{R,\rm f}-h_{R,\rm i})(t/t_{\rm f})^\xi.
\end{equation}
To avoid inserting particles in the bar region, where near-circular orbits
do not exist, particles are not added inside the cutoff radius $R_{\rm cut}$,
which is determined by the current bar length (`adaptive cutoff'; see Paper 1
for details).

\subsection{GMCs}

GMCs are modelled as a population of massive collisionless particles drawn from a mass
function of the form ${\rm d}N/{\rm d}M\propto M^\gamma$ with lower and upper mass 
limits $M_{\rm low}=10^5\msun$ and $M_{\rm up}=10^7\msun$ and an exponent
$\gamma=-1.6$. Their radial density is proportional to the star formation
surface density $\Sigma_{\rm SF}(R)$, and their azimuthal surface density is given by
\begin{equation}
\Sigma_{\rm GMC}(\phi)\propto \left[\Sigma_{\rm ys}(\phi)\right]^\alpha,
\end{equation}
where $\Sigma_{\rm ys}(\phi, R)$ is the surface density of young stars with 
ages 200-400 Myr and $\alpha=1$. The mass in GMCs is determined by the SFR efficiency
 $\zeta$.  Specifically, for each $\Delta m_{\rm stars}$ of stars formed, a total 
GMC mass $\Delta m_{\rm GMC} = \Delta m_{\rm stars}/\zeta$ is created. GMC particles live for
only $50\myr$: for $25\myr$ their masses grow with time as $m\propto t^2$, and for the
final $25\myr$ of their lives their masses are constant, before they disappear
instantaneously. GMCs are added on orbits with $\sigma_0=6\kms$. 
See Paper 1 for more details.

\begin{figure*}
\vspace{-0.cm}
\includegraphics[width=18cm]{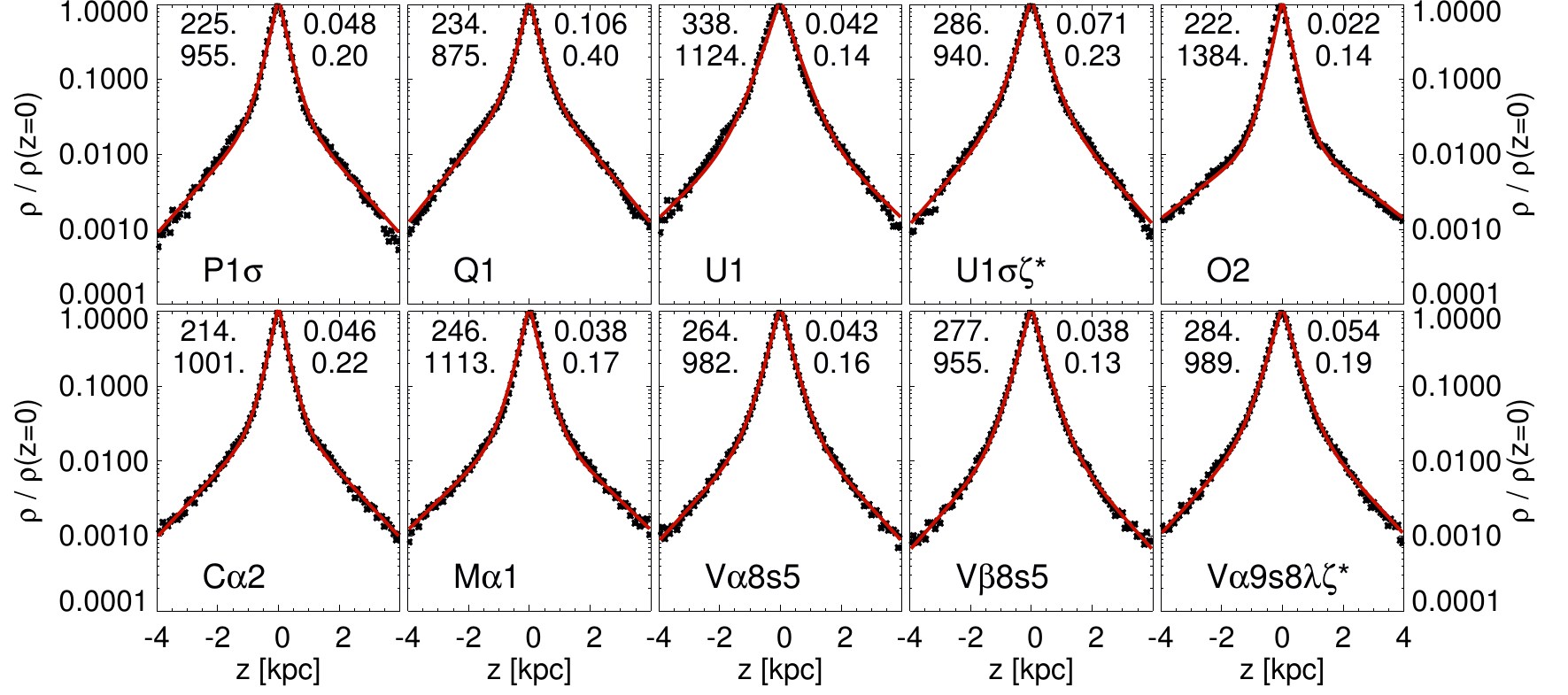}\\
\caption
{Vertical profiles of models at $t=t_{\rm f}$ and $R=8\pm0.5\kpc$ are shown as black points. 
Overplotted are fits of Equation (\ref{eq:vert}) to these profiles. The numbers in the upper-left
corners are the values of the scaleheights $h_{\rm thin}$ and $h_{\rm thick}$ in pc. The 
numbers in the upper-right corners are the values of the density ratio $f$ and the 
surface density ratio $f_{\Sigma}$.}
\label{verts}
\end{figure*}

\subsection{Overview of Models and their Naming}

Table~\ref{modeltable} gives an overview of the models discussed.
We use an extended version of the naming convention described in Paper 1.
All model names start with a capital letter identifying their IC
according to the scheme defined  by Table \ref{ictable}. 

A Greek letter $\alpha$ or $\beta$ following the initial capital letter
indicates that the model has declining birth velocity dispersion $\sigma_0$ rather
than a thick disc in its ICs.  In $\alpha$ models the decline of $\sigma_0$
follows an `atan' shape (Equation \ref{atan}) while in $\beta$ models it
follows a `plaw' shape (Equation \ref{plaw}). For simplicity the model names do not
reflect the specific choices for the parameters of Equations (\ref{atan}) and (\ref{plaw}),
as they are of minor importance for the analyses in this paper.

These capital and, if present, Greek letters are followed by a number between
1 and 9 describing the radial growth history of the model, determined by
parameters $h_{R, {\rm i}}$, $h_{R, {\rm f}}$ and $\xi$. Growth histories
`1', `2' and `5' were already used in Paper 1, `8' ($h_{R, {\rm
i}}=1.0\kpc$, $h_{R, {\rm f}}=4.3\kpc$ and $\xi=0.6$) and `9' ($h_{R, {\rm
i}}=1.0\kpc$, $h_{R, {\rm f}}=3.5\kpc$ and $\xi=0.6$) are new and represent
inside-out growth from a very compact disc into an extended disc.

The final baryonic masses $M_{\rm f}$ of all models lie in the rather narrow
range $5-6\times10^{10}\msun$ and we do not include the variations in the
naming convention. For all other parameters we define standard values and
additional digits added to the model name only when a model deviates in one
or more parameters from the standard. The meanings of the additional digits are:

\begin{itemize}

\item The overall star formation history (SFH) of a model is described by the 
SFR type, the final time $t_{\rm f}$ and the SF time-scale $t_{\rm SFR}$. Our standard choice
is a type 1 SFR with $t_{\rm f}=10 \gyr$ and $t_{\rm SFR}=8\gyr$. We have applied
four additional SFHs, which are labelled by `s5',...,`s8'.

\item The standard input velocity dispersion in models without declining $\sigma_0$ 
is $\sigma_0=6\kms$ . Models with $\sigma_0=10\kms$ are labelled as `$\sigma$'.

\item The GMC star formation efficiency $\zeta$ has a standard value of 0.08.
Models with $\zeta=0.04$ are labelled as `$\zeta$-', and models with $\zeta=0.06$ 
are labelled as `$\zeta$*'.

\item The standard choice for the radial-to-vertical dispersion ratio for inserted 
particles is $\lambda=1$. Models with $\lambda=1.25$ are labelled as `$\lambda$'.

\end{itemize}

\section{Vertical Profiles}
\label{sec:vert}

\citet{juric} studied the vertical stellar density profile in the Snhd and presented a bias-corrected
model fit to Sloan Digital Sky Survey data of the form 
\begin{equation}\label{eq:vert}
\rho(z, R=8\kpc)=\rho_0 \left[ \exp(-|z|/h_{\rm thin})+f\exp(-|z|/h_{\rm thick}) \right].
\end{equation}
They found $h_{\rm thin}=300\pc$ and $h_{\rm thick}=900\pc$ with 
20 per cent uncertainty each and $f=0.12$ with 10 per cent uncertainty.
$f$ here is the ratio of local densities, whereas $f_{\Sigma}=f h_{\rm thick} / h_{\rm thin}=0.36$
is the ratio of local surface densities. \citet{bland} show that while most studies
in the literature find similar scaleheights $h_{\rm thick}$ and $h_{\rm thin}$ to \citet{juric},
this determination ranks at the upper end in terms of $f_{\Sigma}$ for photometric surveys, 
the average literature value of which is significantly lower, $f_{\Sigma}=0.12$.
This strong variation in $f_{\Sigma}$ between different studies is likely caused by significant
differences in the survey selection functions and degeneracies between $f_{\Sigma}$ and 
the scaleheights.

\subsection{Vertical profile shape}

In the models of Paper 1, GMC heating created remarkably exponential
vertical profiles with scaleheights $h_{\rm thin}=200-350\pc$. Thus these
models can reproduce the  scaleheight of the MW's thin disc but fail to create a
realistic thick disc. In Figure \ref{verts} we examine the vertical
profiles of the present models at $t=t_{\rm f}$ and
$R=8\pm0.5\kpc$ as black symbols. We overplot in red fits of Equation
(\ref{eq:vert}) to these profiles with the scaleheights given in the top-left
corner of each panel. The values of $f$ and $f_{\Sigma}$ are given in the top-right
corner of each panel. The figure illustrates that both approaches to
producing thin+thick disc systems presented here produce double-exponential
profiles similar to the one observed in the MW.

In the upper row of Figure \ref{verts} we show profiles for models with thick-disc ICs. 
We find that using disc-like ICs with $z_0\sim1.7\kpc$ yields values of $h_{\rm thick}$ in the 
range $815-1133\pc$ at $t_{\rm f}$ similar to the one inferred for the MW. As was already 
shown in Paper 1, $z_0\sim 1\kpc$ yields final thick discs that are too thin. For elliptical
ICs we find that setting the axis ratio $s=2$ in model K2 yields $h_{\rm thick}\approx2.4\kpc$
and $s=3$ in model O2 yields $h_{\rm thick}\approx1.4\kpc$, which are both too thick. We thus 
decided to focus on disc-like thick ICs. 

In terms of thin-disc scaleheights, we find values $h_{\rm thin}=200-339\pc$. As was already discussed in Paper 1,
for a given GMC mass function, lower values of the star formation efficiency $\zeta\sim0.05$ and thus a 
higher total mass in GMCs per mass of formed stars are required to obtain $h_{\rm thin}\sim 300\pc$ as in the MW.
Lowering the DM halo concentration reduces the vertical force contribution from the halo and thus
also mildly increases $h_{\rm thin}$. The highest values for both $h_{\rm thick}$ and $h_{\rm thin}$ for disc IC models 
are found for model U1 that at $t_{\rm f}$ features an overly extended and thick bar as is discussed
in Section \ref{sec:bar}. As was discussed in Paper 1, bars that extend beyond $R=5\kpc$ can significantly
thicken vertical profiles at $R=8\kpc$.

For the density and surface density ratios in models with thick-disc ICs, we find ranges of $f=0.04-0.167$
and $f_{\Sigma}=0.14-0.53$, which include the \citet{juric} values and are in the upper half of the 
values of the \citet{bland} literature compilation. Naturally, increasing the thick-disc mass $M_{\rm disc, i}$ 
at fixed thick-disc scalelength $h_{R, {\rm disc}}$ increases these ratios (Q versus P models) and decreasing 
$h_{R, {\rm disc}}$ at fixed $M_{\rm disc, i}$ lowers them (U versus Q models). As far as other model parameters
are concerned there are no clear patterns apparent. This is likely connected to competing effects. For example,
few GMCs produce less vertical heating, but lead to stronger bars, which, if long enough, can thicken
the vertical profile.

The lower row of Figure \ref{verts} shows models with declining birth dispersions. We find values for the scaleheights
in the ranges $h_{\rm thin}=214-284\pc$ and $h_{\rm thick}=863-1234\pc$, very similar to the ranges found for
thick IC disc models. Model C$\alpha$2 has the thinnest thin disc as at late times it has a very low value of 
$\sigma_0<5\kms$ and it also has a high star formation efficiency $\zeta=0.08$. \V6 has $\zeta=0.06$ 
and $\sigma_0\sim10\kms$ at late times and thus the highest $h_{\rm thin}$
among these models. V$\alpha$8s5 and V$\beta$8s5 only differ in the shape of the declining $\sigma_0$ curve: 
V$\alpha$8s5 is of type `atan', whereas V$\beta$8s5 is of type `plaw' (see Equations \ref{atan} and \ref{plaw}). 
Their final vertical profiles are rather similar, so both types of decline are acceptable. The value of $h_{\rm thick}$ 
is also mildly influenced by the ICs, which for V and M models is a low-mass $s=2$ elliptical. As models 
V$\alpha$8s5/V$\beta$8s5 have a higher final mass than M$\alpha$1 and thus a lower fraction of IC stars, 
their thick discs appear mildly thinner. 

\begin{figure}
\vspace{-0.cm}
\hspace{-0.2cm}\includegraphics[width=8cm]{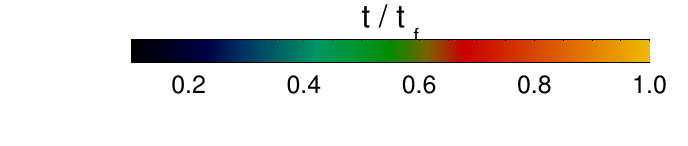}\vspace{-1cm}\\

\includegraphics[width=8cm]{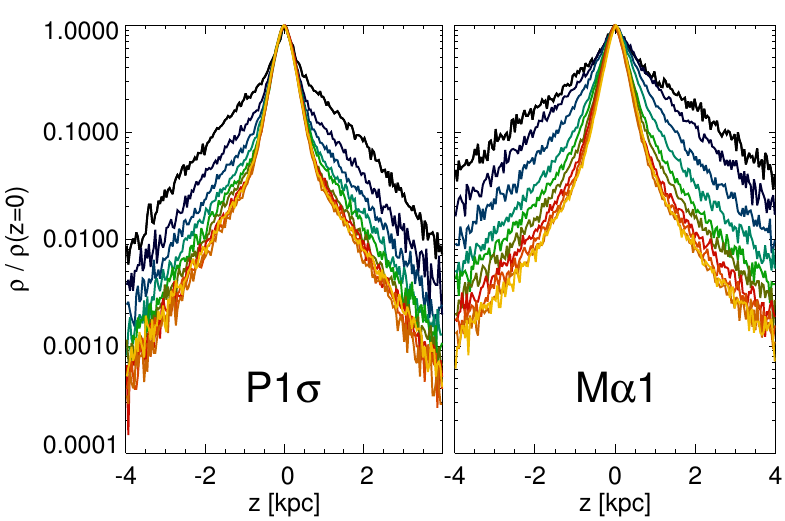}\\
\caption
{Vertical profiles of models at $R=8\pm0.5\kpc$ and various times $t$ as indicated by the colour bar.}
\label{Tverts}
\end{figure}

The corresponding thin-to-thick disc ratios are $f=0.025-0.055$ and $f_{\Sigma}=0.11-0.22$, which are lower 
than the \citet{juric} values, but comfortably within the range found in the literature. For models with declining 
$\sigma_0(t)$ the thick-disc mass fraction depends on the fraction of mass formed during early 
formation stages with high $\sigma_0$ and thus on the detailed forms of $\sigma_0(t)$ and the SFH.
Moreover, the radial growth history determines how many stars are formed at a certain radius during 
this period. Dynamical heating and migration processes also influence the number of old stars found
at $t_{\rm f}$, so the final value of $f$ is not easily predicted.

\subsection{Vertical profile evolution with time}

In Figure \ref{Tverts} we analyse the temporal evolution of the shapes of
vertical profiles in two different types of simulations. In the
models of Paper 1 that included GMC heating, thin-disc vertical profiles are
at all times exponential and their scaleheights change very little with time.
This finding reflects balance between mass growth, which
continuously supplies cold particles and deepens the vertical potential well,
and GMC heating which efficiently increases the vertical velocity dispersions
of young stars.

The thick-disc IC model \P5 shows a double-exponential profile from early on.
The thin-disc part of its profile is very constant, just like the thin-disc-only models of Paper 1.
What changes are the surface density ratio $f_{\Sigma}$, which by construction becomes more
and more thin-disc-dominated, and the scaleheight of the thick disc, which becomes smaller 
with time, because the growth in the thin disc's mass deepens the vertical potential well and the 
thick disc is not heated significantly.

\begin{figure*}
\includegraphics[width=14cm]{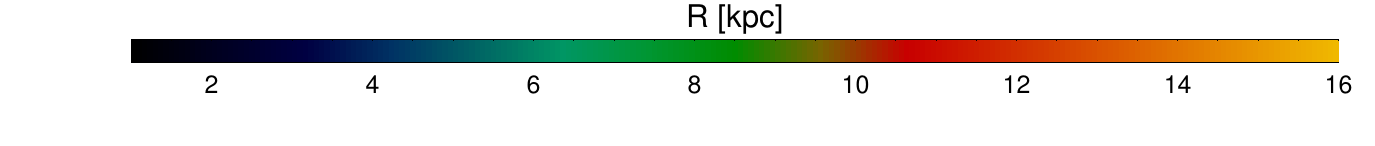}\\
\vspace{-0.5cm}
\includegraphics[width=18cm]{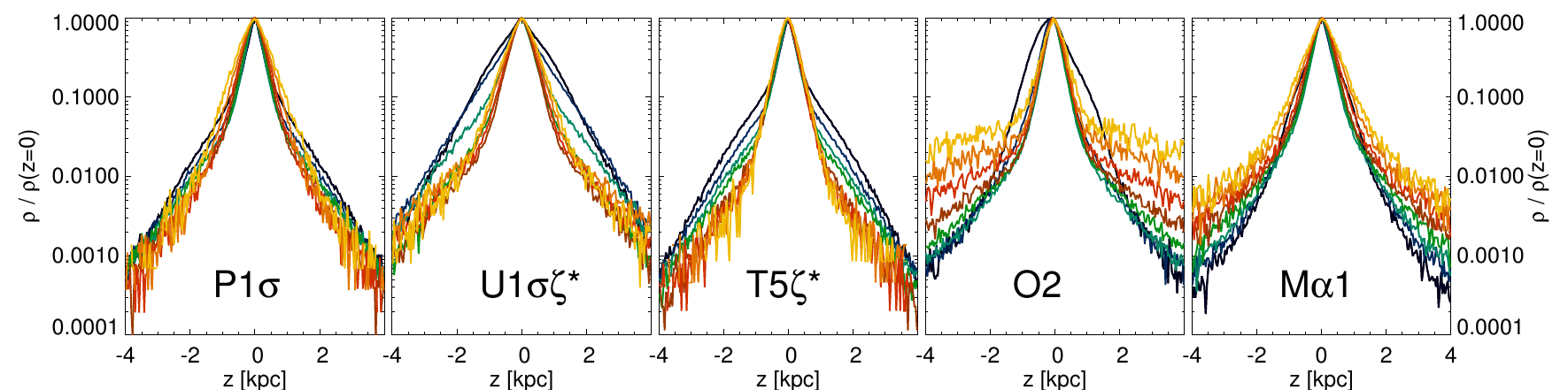}\\
\caption
{Vertical profiles of models at $t=t_{\rm f}$ and various radii $R$ as indicated by the colour bar.}
\label{Rverts}
\end{figure*}

M$\alpha$1 represents models with declining $\sigma_0(t)$. Its profile at early times is closer to a
single- than a double-exponential, as there are no cold, thin-disc populations present.
Only as $\sigma_0(t)$ falls below $\sim 20\kms$ does a thin disc build up. The scaleheight
of the thin disc $h_{\rm thin}$ becomes smaller with time as the decline in $\sigma_0(t)$
cools the thin-disc population as a whole despite the vertical heating due to the GMCs.
The deepening of the vertical potential well adds to that effect and reduces $h_{\rm thick}$
in the same way as in \P5. 

\subsection{Radial dependence of vertical profiles}

The vertical profiles of observed disc galaxies are very constant radially
\citep{vdkruit}. Paper 1 showed that in thin-disc-only models that
include GMC heating, vertical profiles are almost independent of radius $R$,
unless there is a buckled bar, which thickens only the central region.
Figure \ref{Rverts} shows the radial variations of the vertical profiles
in five of our thin+thick models. The inner parts of models \U1o and O2 show
thickening by a bar. \U1o otherwise shows a profile that is almost constant
in radius. 

Model \P5 shows a thin disc that becomes mildly thicker and has a higher mass fraction towards larger $R$,
whereas $h_{\rm thick}$ stays roughly constant. This is caused by inside-out formation, on account of which
the young thin disc has a longer scalelength than the thick disc, and higher $\sigma_0=10\kms$ than
the models studied in Paper 1. Combined with a shallower potential well at outer radii, where
heating is limited, this value of $\sigma_0$ yields thicker thin discs.

\PA1 shows a stronger fading and a mild thickening of the thick populations towards the outskirts. 
Compared to P ICs, the T ICs have a more compact and more massive thick disc and also a bulge component.
The effective vertical profile of the two components thus varies with radius already 
in the ICs. O2 has elliptical ICs, which lead to a thick population that becomes thicker with
increasing $R$ and also attains a higher mass fraction in the outskirts.

Model M$\alpha$1 with declining $\sigma_0$ shows a thin disc that thickens mildly with $R$, because at 
late times it has $\sigma_0\sim10\kms$ and thus behaves similarly to model \P5 discussed above. 
The thick disc becomes thicker with $R$. This is characteristic for models of this type.
It is a consequence of our assumption that $\sigma_0$ is constant with radius. This leads to a
flaring of the hot component, which is hardly influenced by vertical heating. 

Note that in contrast to the thick discs in models with  declining $\sigma_0$, the thick-disc ICs 
of our alternative modelling scheme were set up with a radially constant scaleheight and thus a 
vertical velocity dispersion that declines with $R$.

\begin{figure*}
\hspace{-0.2cm}\includegraphics[width=14cm]{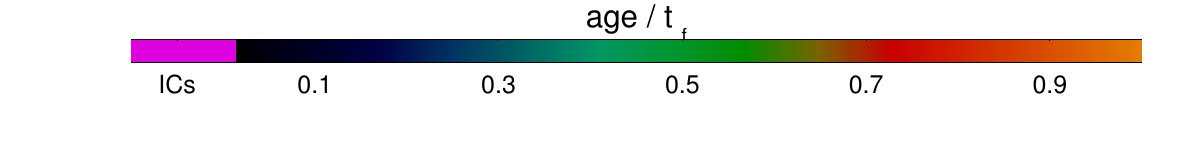}\vspace{-0.4cm}\\

\includegraphics[width=18cm]{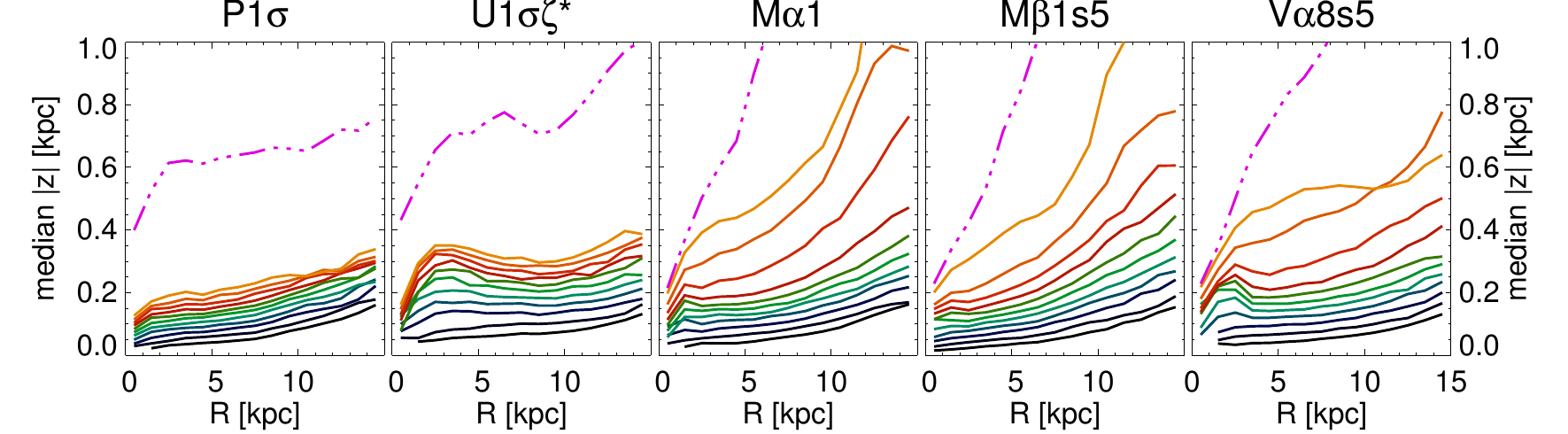}\\
\caption
{The median vertical distance from the midplane $|z|_{\rm med}$ as a function of radius $R$ for
various mono-age populations. Solid lines are for inserted particles of various ages
(see the colour bar). The pink dashed line is for IC stars.}
\label{flares}
\end{figure*}

\subsection{Flaring of mono-age components}

In Figure \ref{flares} we examine for a selection of models how the median
distance from the midplane $|z|_{\rm med}$ varies with radius $R$ for populations of 
different ages. It has been suggested that the lack of significant changes with disc 
radius $R$ in the double-exponential vertical mass profiles of disc galaxies is a 
consequence of inside-out growth combined with $|z|_{\rm med}$ being an increasing 
function of $R$ for all mono-age populations (`flaring'; \citealp{minchev}).

Flaring can be caused by satellite interactions or misaligned infall of gas, which
are not present in our models. However, Paper 1 showed that disc galaxies
formed in isolation
also have flaring mono-age components. The amount of flaring is determined by the 
radial mass profile of the disc and the radial profile of the vertical velocity 
dispersion of mono-age components $\sigma_z(R,\tau)$, which is determined by the birth
dispersions of stars and the vertical heating mechanism(s) at work. Radial migration
of stars can also influence $\sigma_z(R,\tau)$ \citep{sb12}.

The left most panel of Figure \ref{flares} depicts $|z|_{\rm med}(R)$ for the
mono-age populations of model \P5, which has thick-disc ICs. The thin-disc
populations are at all radii substantially thinner than the old thick-disc
stars. These mono-age populations all show flaring, and for the youngest
populations $|z|_{\rm med}$ increases from $R=1$ to $15\kpc$ by up to a
factor of $\eta\equiv|z|_{\rm med}(15\kpc)/|z|_{\rm med}(1\kpc)\sim 5$.  In
fact, in \P5 the structure of the thin-disc populations is similar to that in
thin-disc-only model Y1 examined in Paper 1. The relative increase in
$|z|_{\rm med}$ with $R$ becomes smaller with increasing age, and we find
$\eta\sim 2.5$ for the oldest inserted stars. In model \U1o (second panel in
Figure \ref{flares}) the situation is
altered by a vertically extended bar. On account of bar buckling, $|z|_{\rm
med}$ for intermediate-age populations now peaks around $R\sim3\kpc$, then
declines slightly to $R\sim 8\kpc$, and gradually increases further outward.
The youngest populations have not been affected by bar buckling and have
$\eta\sim 3$.

In all our models, non-IC stars at any given time $t$ are inserted with a radially 
constant birth dispersion $\sigma_0(t)$. On account of the outward decrease in 
surface density, this results in flaring. GMC heating increases $\sigma_z$ more 
strongly in the centre than in the outskirts and thus flattens the increase in 
$|z|_{\rm med}(R)$. For models without buckled bars and with $\zeta=0.04$ and thus 
more GMCs per unit mass of inserted stars, $\eta$ can be as low as $\sim 1.5$ for 
old inserted stars.  Bars can additionally heat the central regions of galaxies 
(see e.g. \citealp{grand}) and thus cause even flatter runs of $|z|_{\rm med}(R)$. 

The three panels on the right of Figure \ref{flares} show models with
declining $\sigma_0$, and in these models the structure of $|z|_{\rm med}(R)$
is quite different from what it is in the models with thick IC discs. Now the
low-mass elliptical ICs are unimportant because the thick disc is formed
mainly by old added stars. Due to the continuous decline in $\sigma_0(t)$,
the curves form a continuum rather than a bimodal grouping. The young,
thin-disc populations behave very similarly to those in the models with
thick-disc ICs, the local maxima in $|z|_{\rm med}$ at $R\sim 2\kpc$ in
model V$\alpha$8s5 being caused by a buckled bar. 

The thick-disc stars in models with declining $\sigma_0$ (red and orange
curves) yield a similar value of $\eta$ to the youngest thin-disc stars in
all models. Again this reflects our decision to make the declining birth
dispersion $\sigma_0$ independent of $R$.  Given that vertical GMC heating
has little influence on the thick-disc populations, and that the depth of the
vertical potential well declines with $R$, strong flaring is an inevitable
consequence. The flattest curves are thus found for the intermediate-age
populations, which were already born on relatively cold orbits and have been
significantly affected by GMC and bar heating. The differences between models
M$\alpha$1 and \M2 can be explained by the different shapes of $\sigma_0(t)$
applied (see Figure \ref{sig0}). Whereas the oldest component of each model
flares in a similar way, the two next-oldest populations show stronger
flaring in the atan model M$\alpha$1 than in the plaw model \M2, in which
$\sigma_0$ declines more gradually.  Also, in the atan model $|z|_{\rm med}$
declines with decreasing age faster than in the plaw model.

The flaring of all disc populations discussed so far can be qualitatively
explained by birth-dispersion profiles and disc heating mechanisms. A
comparison between the thick-disc IC stars in models \P5 and \U1o indicates
that radial migration plays a role as well. The IC stars in \P5 only show a
strong outward increase in $|z|_{\rm med}$ in the centre and hardly any
flaring at larger radii, whereas in \U1o, the flaring of the IC stars in the
outer disc is stronger. The thick-disc ICs were created with a radially
constant scaleheight and are hardly affected by vertical heating mechanisms,
so their curve of $|z|_{\rm med}(R)$ at $t_{\rm f}$ is determined by the
change in the vertical potential well together with extent to which their
stars migrate radially. Bar formation funnels a lot of mass to the centre and
consequently the disc's thickness decreases there. The stronger flaring in
\U1o is likely explained by higher levels of radial migration.

When stars migrate radially, their vertical actions $J_z$ are conserved
\citep{solway}.  Populations of stars that are born with a radially constant
scaleheight in a MW-like disc have their mean $J_z$ decreasing with $R$.
Consequently, outward migrators at a given $R$ have higher $J_z$ (and thus
higher $\sigma_z$) than non- or inward migrators. If there are more outward
than inward migrators, as expected for the outer disc regions, such
populations are expected to flare \citep{sb12}. This principle is complicated by the
finding that stars migrate less if they have high $J_z$ \citep{vera}, but
Paper 4 shows that for our thin-disc populations and for stars from
thick-disc ICs at $t=t_{\rm f}$, outward migrators are indeed more numerous
in the outer disc and have higher $\sigma_z$ than inward migrators (see also
\citealp{roskar}). Model \U1o has both a more compact IC disc and a lower density
dark halo than model \P5, which leads to stronger non-axisymmetries and a higher
fraction of outward migrators in the outer disc (Paper 4), which in turn
explains the stronger flaring of IC stars in \U1o.

\begin{figure*}
\vspace{-0.cm}
\hspace{-0.2cm}\includegraphics[width=14cm]{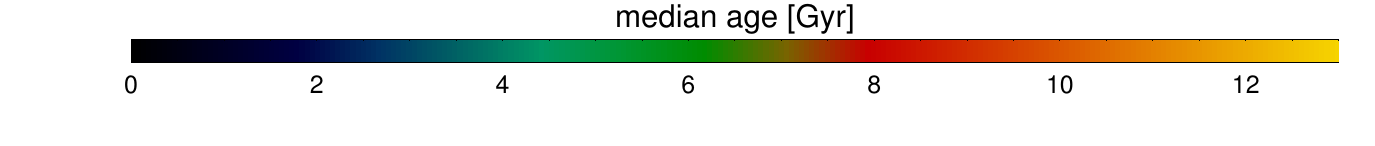}\vspace{-0.6cm}\\

\hspace{-0.5cm}\includegraphics[width=18cm]{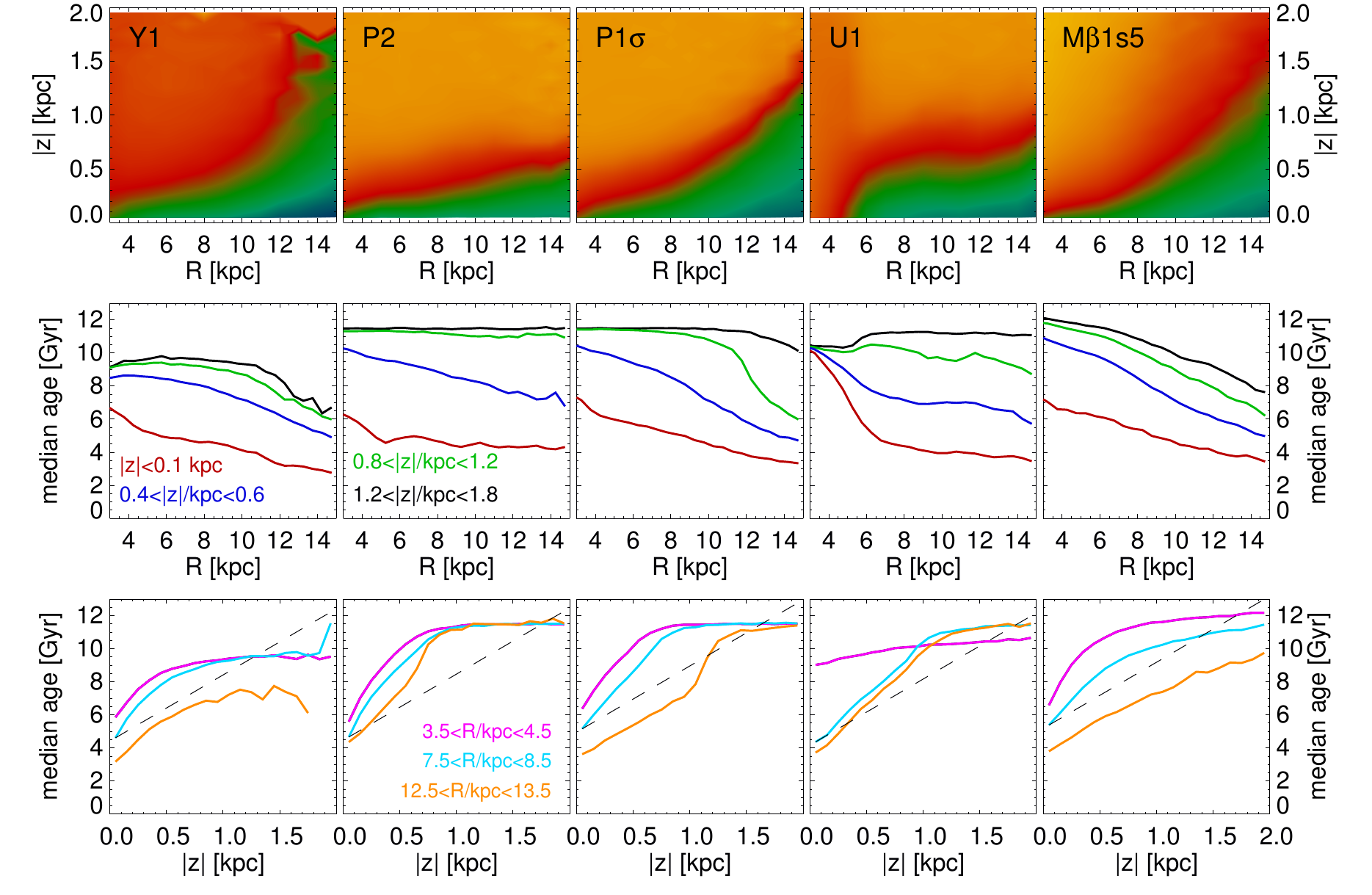}\\
\caption
{Top: median age $\tau_{\rm med}$ maps in the $R$-$|z|$ plane. The colour bar shows the age encoding.
Middle: median ages $\tau_{\rm med}$ as a function of radius $R$ for different vertical distances from the midplane:
red is $|z|<0.1\kpc$, green is $|z|=0.5\pm0.1\kpc$, blue is $|z|=1.0\pm0.2\kpc$ and black is 
$|z|=1.5\pm0.3\kpc$. Lower: median ages $\tau_{\rm med}$ as a function of vertical distance from the midplane $|z|$ at 
different radii: pink is $R=4\pm0.5\kpc$, cyan is $R=8\pm0.5\kpc$, orange is $R=13\pm0.5\kpc$. The dashed line shows
a constant vertical gradient of $4\gyr/\kpc$. Each column presents one model.}
\label{agezr}
\end{figure*}

Compared to the oldest inserted stars in models M$\alpha$1 and \M2, in model
V$\alpha$8s5 this population shows a much flatter curve $|z|_{\rm med}(R)$ at
$R>5\kpc$. Model V$\alpha$8s5 has a lower density dark halo and a more
compact disc at early times than the M models. Hence in this model the oldest
inserted stars in the outer disc have a higher fraction of outward
migrators.  For models with declining $\sigma_0$, thick-disc stars are born
with radially constant $\sigma_z$ and thus their mean $J_z$ increasing with
$R$. As they are hardly affected by disc heating, in these models old
outward migrators at a given $R$ have lower $J_z$ and thus lower $\sigma_z$
than inward migrators. This likely explains the flatter curve $|z|_{\rm
med}(R)$ for the oldest inserted stars in model V$\alpha$8s5. So depending on
the shape of $\sigma_z(R)$ at birth, radial migration can both strengthen and
weaken the flaring of a mono-age population. A detailed analysis of radial
migration in our models is presented in Paper 4.

\subsection{Radial and vertical age structure}

The combination of recent and ongoing astrometric and spectroscopic surveys of MW stars is about
to increase vastly our knowledge of the age structure in the Galactic disc(s) (e.g. \citealp{martig}).
In the top row of Figure \ref{agezr} we therefore show maps of median age $\tau_{\rm med}$ as a function of 
$R$ and $|z|$ for various models. To do so we assume that the oldest stars in all models are $13\gyr$
old and thus randomly assign ages in the range $[t_f,13\gyr]$ to IC star particles.

The leftmost panel shows the thin-disc-only model Y1 from Paper 1. The innermost region $R<4\kpc$ 
at all altitudes and high altitudes $|z|>1.5\kpc$ at all radii are dominated by old stars. The youngest
$\tau_{\rm med}$ are found at high $R$ and low $|z|$. Intermediate $\tau_{\rm med}$ are confined to $|z|<200\pc$
at $R<6\kpc$, to $|z|<500\pc$ at $R<10\kpc$, but can populate regions up to $|z|\sim1500\pc$ at $R\sim15\kpc$.
This characteristic $\tau_{\rm med}$ pattern is caused by a combination of inside-out formation and
disc heating. The oldest population is more compact and thicker, whereas the younger population
lives closer to the midplane and preferentially at larger radii, where it extends to higher $|z|$.

Model \P5 has an old thick disc, which is more massive, much thicker and more extended 
than the oldest disc population in Y1. This changes the $\tau_{\rm med}$ map only marginally. The high-$|z|$
populations are by construction older than in Y1 and the young, outermost populations
are also mildly older than their counterparts in Y1. The $\tau_{\rm med}$ pattern is however rather similar. 

P2 differs from \P5 in lacking inside-out formation, which makes the outer populations older. It also has lower 
$\sigma_0$, which reduces the flaring of the outer populations. Consequently, at all radii $R$, all altitudes 
$|z|>500\pc$ are dominated by old stars and the zone populated by intermediate $\tau_{\rm med}$ is confined 
to $|z|>300\pc$ at $R<10\kpc$ and to $|z|>500\pc$ at all $R$. 

\begin{figure*}
\vspace{-0.cm}
\hspace{-0.5cm}\includegraphics[width=16cm]{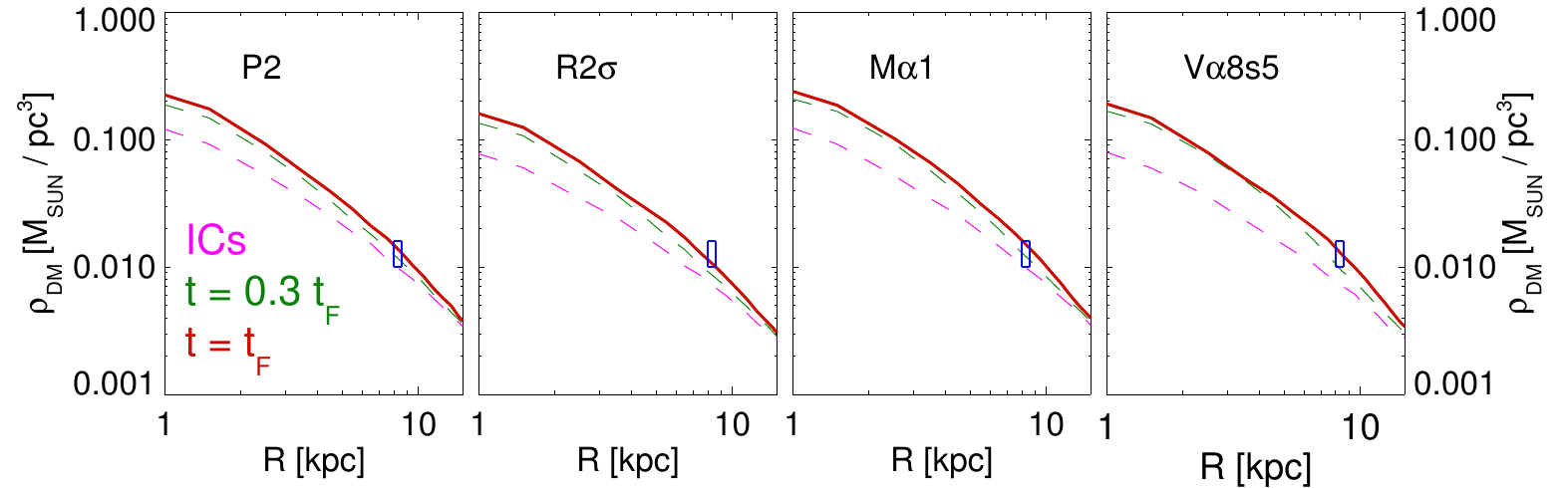}\\
\caption
{DM density profiles $\rho_{\rm DM}(R)$ as measured in the midplane of the galaxies P2, \R2, M$\alpha$1 and V$\alpha$8s5.
The solid red lines show the models at $t=t_{\rm f}$, whereas the dashed lines show $t=0.3t_{\rm f}$ (green) and the ICs 
(pink). The blue boxes mark the constraints for today's Snhd from \citet{mckee}.}
\label{halo}
\end{figure*}

Model U1 is an inside-out model, but compared to \P5 has a lower concentration halo, a more massive
and more compact IC thick disc and lower $\sigma_0$. It has a thicker and longer bar than the other
depicted models. The buckled bar causes an area at $R<5\kpc$, which is populated by stars with old
$\tau_{\rm med}$ and shows no noticeable vertical age gradient. The bar
also heats the disc at $R\sim5-10\kpc$ vertically and thus increases the altitudes at which younger 
stars are found, so the radial increase of the maximum $|z|$ at which intermediate $\tau_{\rm med}$
are found is flatter in this model.

Model \M2 is a model with declining $\sigma_0$. It shows a $\tau_{\rm med}$ pattern that differs from those of the other models
shown in Figure \ref{agezr}. At $R<5\kpc$, its vertical age structure is similar to that in the P models,
but at $R>5\kpc$ the intermediate-age stars reach higher altitudes, which leads to declining radial age gradients at all $|z|$.
This is caused by our assumption of radially constant $\sigma_0(t)$, as was also discussed in relation to the
flaring of mono-age populations shown in Figure \ref{flares}. Moreover, vertical heating is inefficient for
stars with high $\sigma_0$ and for stars at large radii.

\citet{martig2} have recently presented measurements of radial gradients in $\tau_{\rm med}$
at various altitudes $|z|$ in the Snhd. At $|z|>500 \pc$, they find
significant declines in $\tau_{\rm med}$ with radius $R$ at all radii.
However, the value of $\partial \tau_{\rm med} / \partial R$ is still very uncertain.
Close to the plane near the solar radius $R_0$, $\tau_{\rm med}(R)$ is rather flat. In
the middle row of Figure \ref{agezr} we show $\tau_{\rm med}$ versus $R$ at four
ranges in $|z|$.

As model Y1 lacks a thick disc, we will not discuss it in detail. The P and U models with thick IC discs
clearly show no decline in $\tau_{\rm med}$ with $R$ at $|z|\sim 1.5\kpc$ and only \P5 shows a negative $\rd \tau_{\rm med} / \rd R$
at $|z|\sim 1.0\kpc$ and $R>10\kpc$ due to inside-out formation and a stronger flaring in the young disc due 
to higher $\sigma_0$. By contrast, \M2 shows clear negative gradients throughout the whole disc at all latitudes and is thus
more in agreement with the age determinations of \citet{martig2} .

At $|z|\sim 0.5\kpc$, all models show a negative radial $\tau_{\rm med}$ gradient. This gradient is weaker in U1 due
to the influence of the unrealistically long and thick bar. In the midplane, the inside-out models also show clear
negative age gradients, whereas P2, which has a constant radial feeding scalelength, shows a rather
flat age profile, as does U1, again strongly affected by the long bar. As the age gradients in \M2 and \P5
are flatter closer to the midplane and the observations are still very uncertain, little can be deduced
yet about inside-out growth.

\citet{casagrande} have presented evidence for a vertical age gradient in the Snhd from asteroseismology.
They find a decline by $\sim4\pm 2\gyr/\kpc$, but have little knowledge of the shape of the decline.
In the lower row of Figure \ref{agezr} we present $\tau_{\rm med}$ as a function of $|z|$ at three radii,
the cyan line representing a solar-like radius and the dashed line showing a constant vertical gradient
of $4\gyr/\kpc$.

Due to the dominance of the thick disc at $|z|>1\kpc$ and the weak radial age gradient, at all radii
in P2, age increases more strongly than at $4\gyr/\kpc$ up to $|z|\sim 1\kpc$ and then flattens
out. In \P5 at $R=13\kpc$, the enhanced presence of younger stars away from the plane results in a 
flatter increase of $\tau_{\rm med}$ with $|z|$, but at $R=8\kpc$ the situation is similar to that in P2. In U1
the long bar causes an almost flat $\tau_{\rm med}$ versus $|z|$ plot in the bar region and a flatter gradient at $R=8\kpc$.
In \M2, $\tau_{\rm med}(|z|)$ at $R=8\kpc$ flattens more gradually and at $R=13\kpc$ has an almost constant slope
due to the stronger flaring of mono-age populations.

Averaged over the studied vertical extent of $2\kpc$, all models show a vertical gradient consistent with the
findings of \citet{casagrande}. As they observe very few stars above $|z|>1.2\kpc$, model \M2 shows the best
agreement with the still very uncertain data. 

\section{Radial mass distribution}
\label{sec:rad}

In this section we investigate the radial distribution of baryonic and dark matter in our models
and the circular speed curves $v_{\rm circ}(R)$ that result from them.

\subsection{Dark matter density}

As discussed in Paper 1 the parameters for our DM haloes as set up in the ICs are motivated
by what $\Lambda$ cold dark matter ($\Lambda$CDM) predicts for haloes associated with MW mass galaxies. The DM profile
$\rho_{\rm DM}(r)$ will be modified by growing a massive baryonic disc within the DM halo and
by interaction with non-axisymmetric disc structures such as the bar and spirals. As the
halo is always spherical in the ICs, but the disc mass fraction in the ICs varies strongly
between ICs and the various galaxy models evolve differently, the final haloes differ
even if two models share the same DM IC parameters.

\begin{figure*}
\vspace{-0.cm}
\hspace{-0.5cm}\includegraphics[width=18cm]{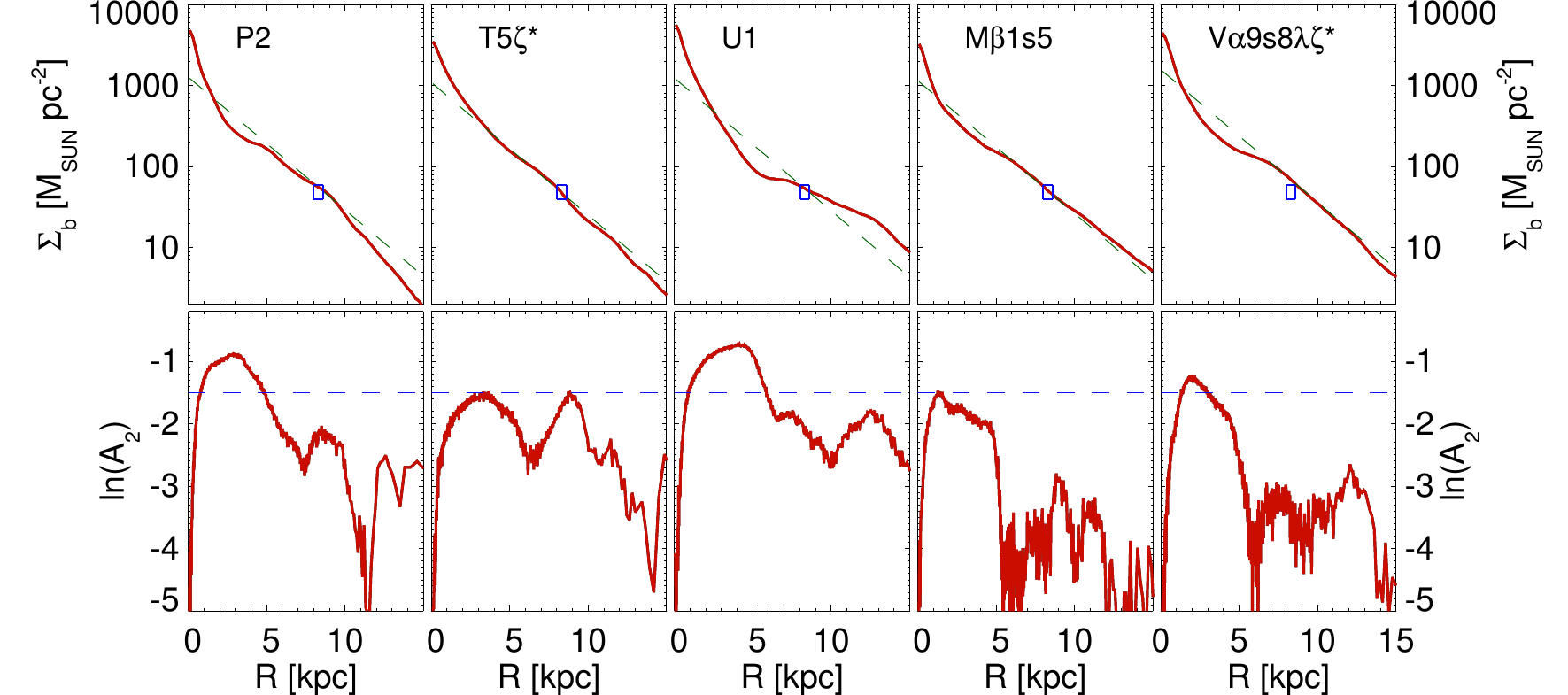}\\
\caption
{Top: baryonic surface density profiles $\Sigma_{\rm b}(R)$ for various models. 
The blue boxes mark the constraints for the Snhd from \citet{mckee}.
The dashed line marks an exponential with scalelength $h_R=2.65\kpc$.
Bottom: radial profiles of the $m=2$ Fourier amplitude $A_2$. The dashed lines
mark $\ln(A_2)=-1.5$.}
\label{sd}
\end{figure*}

Our best constraints on the DM content in the MW come from dynamical measurements
of the total matter surface density in the Snhd. Subtracting the baryonic components, \citet{mckee}
find a local DM density of $\rho_{\rm DM}=0.013\pm0.003\msun \pc^{-3}$. In Figure \ref{halo}
we plot in red the DM density profiles $\rho_{\rm DM}(R)$ as measured in the midplane of the
galaxy at $t=t_{\rm f}$ and compare them to the Snhd constraints assuming that the solar
Galactic radius is $R_0=8.3\pm0.3\kpc$ \citep{ralph2012}.

We find that  at $t_{\rm f}$ all models fall within the constraints of \citet{mckee}. The four shown in 
Figure \ref{halo} are a representative selection. The main drivers for $\rho_{\rm DM}(R_0)$
at $t_{\rm f}$ are, as expected, initial halo concentration $c$ and the added disc mass 
$M_{\rm add}=M_{\rm f}-M_{\rm disc, i}-M_{\rm bulge, i}$. Consequently \R2, which has $c=6.5$ and  
$M_{\rm add}=3.5\times10^{10}\msun$ has the lowest $\rho_{\rm DM}(R_0)$, whereas M$\alpha$1 with
$c=9$ and $M_{\rm add}=4.5\times10^{10}\msun$ has the highest $\rho_{\rm DM}(R_0)$ among the 
models shown. P2 with $c=9$ and $M_{\rm add}=3.5\times10^{10}\msun$ and V$\alpha$8s5 with 
P2 with $c=6.5$ and $M_{\rm add}=5.5\times10^{10}\msun$ show intermediate $\rho_{\rm DM}(R_0)$.
It is worth noting that the models with $c=4$ presented in Paper 1 and discarded because
of overly strong bars indeed show too low $\rho_{\rm DM}(R_0)$.

We also note that at $t_{\rm f}$ none of our models shows a cored DM profile in the centre, as was recently
favoured by \citet{cole}. Our IC DM profiles do not contain a core as is indicated by the pink
dashed lines. The DM densities $\rho_{\rm DM}(R_0)$ of the ICs are significantly lower than 
in the final models. Initial profiles with $c=9$ lie at the lower allowed limit for today's Snhd and
models with $c=6.5$ are clearly below this limit. During the simulations they are altered by 
compression due to the added mass in stars and by angular momentum transfer from stars to 
DM due to bars and spirals. As all models have declining ${\rm SFR}(t)$ and bars form in the
later evolution stages as shown in Section \ref{sec:bar}, the increase in the DM density
at $R<15\kpc$, where the disc grows, is strong up to $t=0.3t_{\rm f}$ as indicated by the
green dashed lines and rather weak afterwards. In the four models shown, the relative increase
in $\rho_{\rm DM}(R<15\kpc)$ is strongest in models V$\alpha$8s5 as it has the largest
baryonic mass fraction and the largest $M_{\rm add}$, and the increase is weakest in P2. Angular momentum transfer
to the halo by spirals and the bar is not strong enough to create cores, as was also 
shown by \citet{sellwood08}.

\subsection{Solar Neighbourhood surface density}
\label{sec:sd}

Paper 1 showed that despite having control over the evolution of the input scalelength
$h_R (t)$, there was little control over the final surface density profile of the models. The more compactly
a disc was fed, the earlier it grew a bar, which redistributed matter and,
as we avoid inserting particles into the bar region, shifted the inner cutoff radius outwards.
In the end, the surface density profiles of a range of models with different radial growth histories
were thus rather similar. Our surface density was thus decided by the total mass of the final model,
which we justified from the reasonable agreement of our models with a) the vertical scaleheight 
of the thin disc, b) an appropriate local circular speed, c) an appropriate amount
of radial migration to $R_0$ and d) appropriate vertical and radial velocity dispersions.

In Figure \ref{sd} we examine how well our models fulfil constraints on the Snhd baryonic surface 
density $\Sigma_{\rm b}(R_0)$. Table 3 of \citet{mckee} gives an overview of determinations of 
$\Sigma_{\rm b}(R_0)$ including gas of all phases, stars and stellar remnants. $\Sigma_{\rm b}$
is consistently found to be in the range $\Sigma_{\rm b}(R_0)=40-60\msun\pc^{-2}$, where the given errors
are included in the interval. Gas is found to contribute $25-30$ per cent. As in our simulations
gas is only represented by GMCs, our gas fractions are much lower. 

\begin{figure*}
\includegraphics[width=14cm]{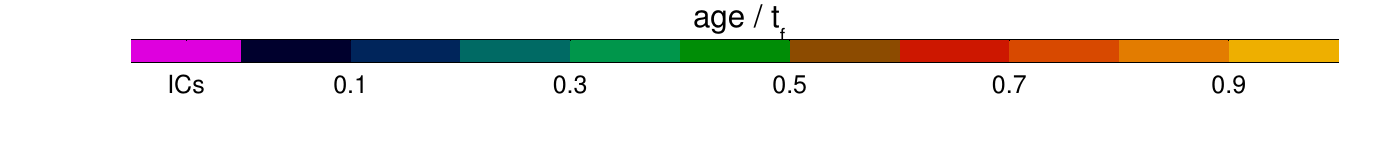}\\
\vspace{-0.7cm}
\hspace{-0.5cm}\includegraphics[width=18cm]{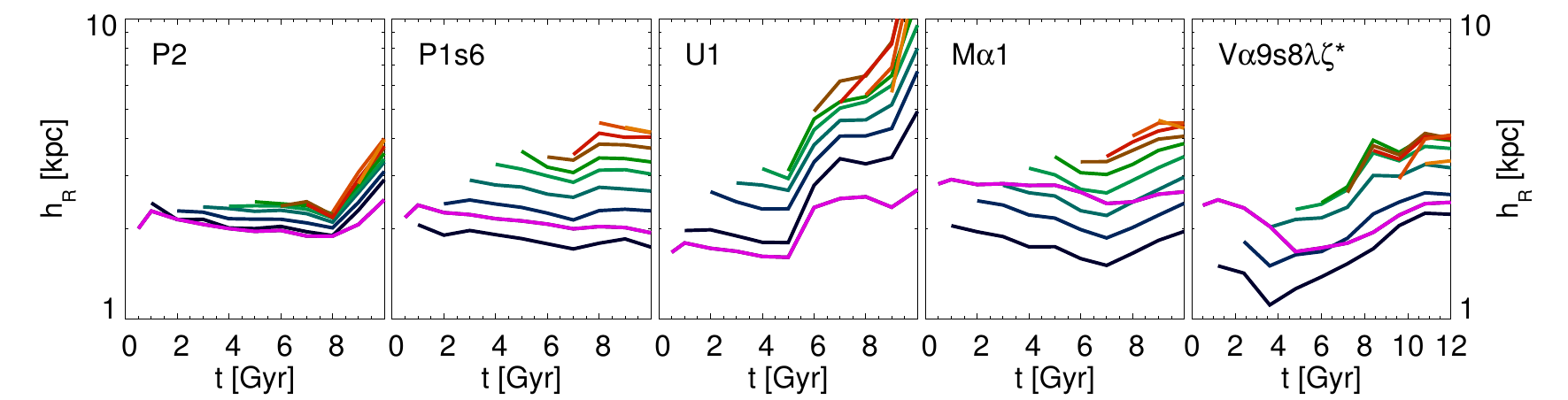}\\
\caption
{Evolution with time of exponential scalelength $h_R$ at $R=8\kpc$ for various mono-age components.}
\label{hr}
\end{figure*}

We choose $\Sigma_{\rm b}(R)$ over the stellar surface density $\Sigma_{\star}(R)$, as we are interested in the 
connection between disc structure and kinematics, and the strength of non-axisymmetries and thus the levels of radial
disc heating and radial migration are determined by $\Sigma_{\rm b}(R)$. Moreover, the interplay between vertical
profiles and vertical velocity dispersions depends on the total mass surface density and not only on $\Sigma_{\star}(R)$ 
and it is therefore appropriate to compare $\Sigma_{\rm b}$ in models and observations, although the division 
of mass between gas and stars is very different. It should be noted that the neutral hydrogen component, 
which is missing in our models, will have a smaller scaleheight than the stars and thus a model that has the 
right vertical profile, kinematics and DM halo is not expected to agree with the Snhd $\Sigma_{\rm b}(R_0)$.

The local radial exponential scalelength $h_R (R_0)$ of the MW is rather uncertain. \citet{licquia}
recently compiled a variety of measurements in the optical and infrared, the vast majority of which
fall in the range $2-4\kpc$. Their meta-analysis of 29 previous measurements yields an estimate
of $h_R(R_0)\sim2.65\kpc$. \citet{bovy} showed that populations of stars with different chemical abundances
show widely varying scalelength, the most compact of which have $h_R\sim 1.5 \kpc$ and the most
extended of which are consistent with locally flat profiles. As we are plotting $\Sigma_{\rm b}(R_0)$
that includes GMCs, the comparison is not exact, but because at $t=t_{\rm f}$ the GMC mass fractions
are $2-3$ per cent as for molecular gas in the MW today, the correction is negligible for our purposes.
A more relevant question is whether the missing neutral gas mass, which is a highly relevant
mass component in the outer MW disc, is properly represented in our models. 

In the upper row of Figure \ref{sd} we plot $\Sigma_{\rm b}(R)$ for various models and overplot a blue box indicating
$\Sigma_{\rm b}(R_0)=40-60\msun\pc^{-2}$ at $R_0=8.3\pm0.3\kpc$ and a dashed line indicating an exponential
with $h_R (R_0)=2.65\kpc$. Due to the connection of $\Sigma_{\rm b}(R)$ to bars, we also plot the $m=2$ Fourier 
amplitude
\begin{equation}
 A_2(R)\equiv{{1}\over{N(R)}} \sum\limits_{j=1}^{N(R)}e^{2\i\phi_j}
\label{a2}
\end{equation}
in the lower row of Figure \ref{sd}. The dashed line marks $\ln{A_2}=-1.5$, which is used
for determining the adaptive cutoff region, within which no particles are inserted in our models. 

Model P2 has a constant feeding scalelength $h_R=2.5\kpc$, an IC disc scalelength 
$h_{R, {\rm disc}}=2.5\kpc$ and lives in a $c=9$ halo. Despite the constant input scalelength the final profile
is very different from a simple exponential. At $R<5\kpc$, the profile is shaped by the bar, 
which at $t_f$ has a length of $\sim 5\kpc$, similar to that of the MW bar. The bar steepens the
profile in the centre and flattens it at radii similar to those of the bar tips. At $R=5-10\kpc$
the surface density profile is mildly flatter than the dashed $h_R (R_0)=2.65\kpc$ line, whereas
at $R>10\kpc$ the profile is steeper. The Snhd surface density is close to the upper limit of the 
observed range $=40-60\msun\pc^{-2}$.

Model \PA1 has a more compact and more massive IC thick disc than P2, inside-out formation in the range $1.5-3.5\kpc$,
a mildly higher final mass and a higher GMC mass fraction. Its thick disc also has a higher-than-average
ratio of radial to vertical velocity dispersions $\sigma_R^2/\sigma_z^2=1.8$. The outcome is a model with 
a weaker bar and thus a $\Sigma_{\rm b}(R)$ profile that is well-fit by an exponential at $R=3-15\kpc$. $\Sigma_{\rm b}(R_0)$ 
agrees well with the Snhd constraints. This is one of the models which comes closest to the inferred local profile of the MW.

Model \M2 has declining $\sigma_0$. It lives in a $c=9$ halo and has inside-out growth in the range 
$1.5-4.5\kpc$, which generates a final exponential mildly flatter than $h_R (R_0)=2.65\kpc$. Its value of
$\Sigma_{\rm b}(R_0)$ is in agreement with the Snhd constraints. It has a rather weak bar, which influences
the profile only at the inner radii. 

Model U1 lives in a $c=7.5$ halo and has a massive and compact thick IC disc and inside-out formation in the range 
$1.5-4.5\kpc$. It has a stronger and longer bar compared to the two previous models with $c=9$ haloes
due to a higher baryon fraction as discussed in Paper 1. Its profile shows a steep bar region out to $R\sim5\kpc$,
a flat region at $R\sim5-8\kpc$ and a shallow exponential decline at $R>8\kpc$. Like for the previous models,
 $\Sigma_{\rm b}(R_0)$ agrees well with Snhd constraints. 

The declining $\sigma_0$ model \V6 has a $c=6.5$ halo, a high final mass $M_{\rm f}=6\times10^{10}\msun$ and 
grows inside out in the range $1.0-3.5\kpc$. Its $\Sigma_{\rm b}(R_0)$ is too high for MW constraints and 
although its bar is weaker than in U1 and also P2, its $\Sigma_{\rm b}(R)$ profile is significantly flattened
at $R\sim4-7\kpc$. At $R\sim7-12\kpc$ the profile agrees well with an exponential with $h_R (R_0)=2.65\kpc$.

There is a clear tendency of $A_2$ being lower in models with declining $\sigma_0$ compared to models with 
thick IC discs. We will discuss this further in Section \ref{sec:bar}.

\subsection{Radial profile evolution of mono-age components}

\begin{figure*}
\vspace{-0.cm}
\hspace{-0.5cm}\includegraphics[width=18cm]{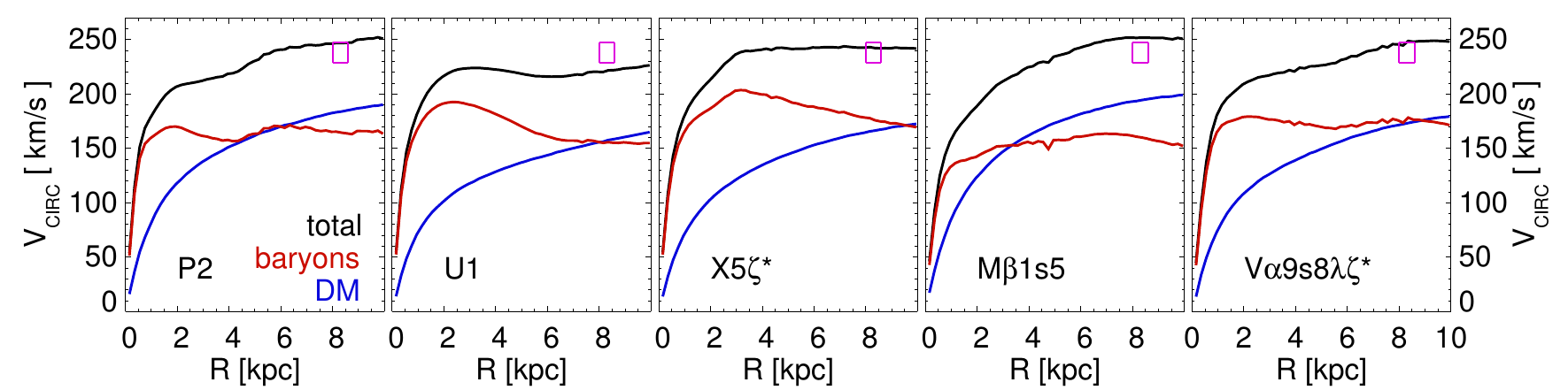}\\
\caption
{Circular speed curves $v_{\rm circ}(R)$ measured in the midplane of the disc and averaged azimuthally.
Blue lines mark the contributions from DM and red lines the baryonic contribution. Pink boxes mark
the constraints on $R_0$ and $v_{\rm circ}(R_0)$ determined by \citet{ralph2012}.}
\label{vcirc}
\end{figure*}

Paper 1 and Section \ref{sec:sd} have shown that the output scalelengths are somewhat independent
of the input scalelengths, as bars and spirals redistribute matter. We know from observations of stars
in the Snhd that the old thick populations are more compact than the young thin ones \citep{bovy}.
To test, how different populations of stars are affected by changes in the radial distribution, in 
Figure \ref{hr} we plot local scalelengths $h_R(R_0)$ as a function of time for populations of different
ages. The pink line is for IC stars and the other colours are for 10 equally spaced age bins of all stars
inserted during the simulations. $h_R(R_0)$ is determined by a single-exponential fit to the
surface density profile $\Sigma(R)$ at $R=6-10\kpc$, irrespective of how good the fit is.

Model P1s6 has little bar activity at any stage of its evolution and is thus well
suited to understand the plots. As it has inside-out growth from $1.5$ to $4.3\kpc$ paired with a thick IC disc
with $h_{R, {\rm disc}}=2.5\kpc$, different age components are rather well separated
in size. The IC component has a shorter scalelength than in the setup, because it is compressed by
the disc's gravitational field. Apart from a mild shrinking of all populations due to compression
and a mild level of noise, which is likely caused by spiral activity, the output scalelengths
are essentially set by the input scalelengths.

P2 has a constant input scalelength $h_R=2.5\kpc$ and the same ICs as P1s6 and thus the populations of
different ages are only mildly separated in size due to continuous compression. At $t\sim 8\gyr$ bar
formation causes an increase in $h_R(R_0)$, which is stronger for younger populations, so the measured
scalelengths increase to $h_R(R_0)\sim2.5\kpc$ for the oldest and $\sim 4\kpc$ for the youngest components. As already 
shown in Figure \ref{sd}, U1 is more strongly affected by a bar. U1 has inside-out growth as in P1s6
and a shorter IC disc scalelength $h_{R, {\rm disc}}=2.0\kpc$. Bar formation at $t\sim4\gyr$ causes a strong 
increase in $h_R(R_0)$ for all age groups and at $t\sim9\gyr$  bar growth causes another increase, 
so at $t=t_f$, the oldest population has $h_R(R_0)\sim2.5\kpc$ and the youngest populations have an
essentially flat profile. 

The declining $\sigma_0$ models M$\alpha$1 and \V6 have low-mass elliptical ICs, for which we find $h_R(R_0)\sim2.5\kpc$
fits at early times. They both grow inside out, M$\alpha$1 from $1.5$ to $4.3\kpc$ and \V6 from $1.0$ to $3.5\kpc$.
As the initial mass of the baryonic ICs is much lower than in P and U models, the amount of compression
for the oldest components is stronger. This is especially true for \V6, which has more compact feeding scalelengths
at early times and also a shorter-than-average SFR time-scale $t_{\rm SFR}=6\gyr$. The latter increases the mass in 
stars added at early times and thus also the mass in GMCs present at these formation stages.
M$\alpha$1 has a bar from $t\sim 6\gyr$, which causes a mild increase for all $h_R(R_0)$,
whereas bar activity is measurable from $t\sim 3\gyr$ onwards in \V6.

Irrespective of how long a bar is and how strongly it affects disc evolution and how high is the level of compression,
the final ordering of $h_R(R_0)$ always reflects the ordering of scalelengths at input. Additionally, in combination
with the results of Section \ref{sec:bar}, it is clear that all models with a bar similar to that of the MW
show an increase for $h_R(R_0)$ of all age components with time due to bar formation and growth.

\subsection{Circular speed curves}
\label{sec:vc}

Recently, various surveys of bulge/bar stars and microlensing data have enabled
more detailed mass models of the centre of the MW \citep{wegg, portail, cole}.
These models agree in the following points: 1) The centre of the MW is baryon
dominated; 2) The baryonic contribution to the rotation curve at $R\sim3\kpc$
is $v_{\rm circ, b}\sim185\kms$ [although \citealp{wegg} find an uncertainty
$\sim \pm 25 \kms$]; 3) The contributions of DM and baryons to $v_{\rm circ}$ are
roughly equal at $R_0$. Further constraints on $v_{\rm circ}(R)$ come from the
motion of stars in the Snhd: \citet{ralph2012} finds $v_{\rm circ}(R_0)=238\pm9\kms$
and $R_0=8.3\pm0.3\kpc$.

\citet{as15} (hereafter AS15) presented an inside-out growing model in a $c=9$ halo. Its rotation
curve fulfilled the \citet{ralph2012} constraints, but has too few baryons in the centre
to match any of the constraints from microlensing. In Figure \ref{vcirc} we present
circular speed curves for a selection of our models: black is total $v_{\rm circ}(R)$,
red is the baryonic contribution $v_{\rm circ, b}(R)$ and blue is the DM contribution
$v_{\rm circ, DM}(R)$. The pink boxes mark the \citet{ralph2012} constraints.
$v_{\rm circ}(R)$ is measured in the midplane of the disc and averaged azimuthally.

Figure \ref{vcirc} shows two $c=9$ models: P2 and \M2. P2 has $h_R=2.5\kpc$ both in the
ICs and at all times through the simulation. Its value of $v_{\rm circ}(R_0)$ is at the upper end
of allowed values. Its central baryonic contribution is higher than that of AS15
but still too low for the microlensing constraints. \M2 grows inside out
in the range $1.5-4.3\kpc$ and has a higher added mass $M_{\rm add}$. The former
leads to a weaker central baryonic contribution than in P2 and the latter causes
a stronger compression of the halo and thus an unacceptably high $v_{\rm circ}(R_0)$.

\begin{figure*}
\vspace{-0.cm}
\hspace{0.cm}\includegraphics[width=18 cm]{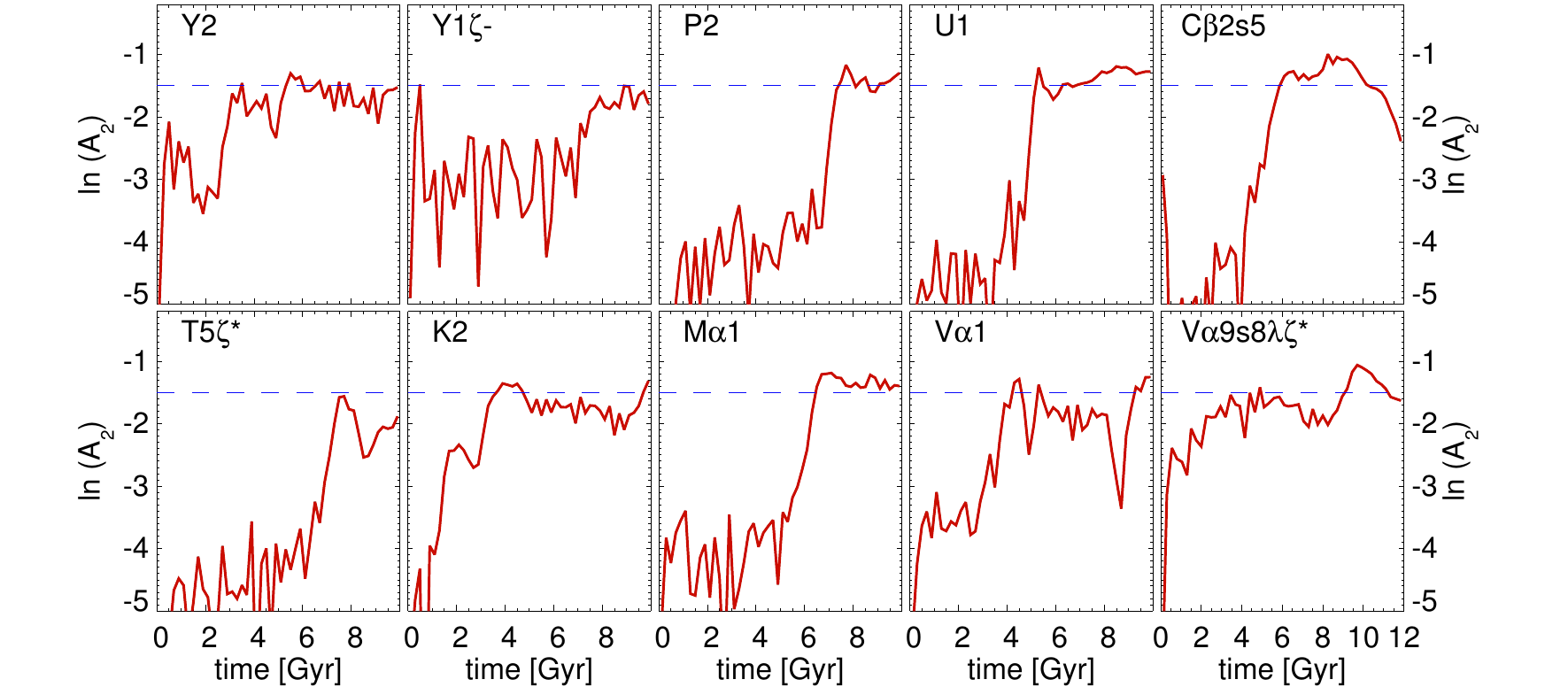}\\
\caption {Evolution with time of the $m=2$ Fourier amplitude $A_2$ measured
within $R=3\kpc$. The blue dashed lines mark $\ln(A_2)=-1.5$. Note that of the shown models
only \V6 and \C3b have $t_{\rm f}=12\gyr$ and all others have $t_{\rm f}=10\gyr$.} \label{barx}
\end{figure*}

The microlensing constraints suggest shifting mass from the halo to the discs.
As was shown above, the Snhd DM density allows concentrations as low as $c=6.5$ given
a constant IC halo mass of $M_{\rm tot}=10^{12}\msun$. Models U1 and \WA1 have haloes with
$c=7.5$ and \V6 has a $c=6.5$ halo. U1 has a massive thick IC disc with $M_{\rm disc,i}=2.5\times10^{10}\msun$
and $h_{R, {\rm disc}}=2.0\kpc$. It grows inside out from $1.5$ to $4.3\kpc$ reaching a final
mass of $M_{\rm f}=6\times10^{10}\msun$. Its baryonic contribution to the rotation curve
peaks at $v_{\rm circ, b}\sim190\kms$ and falls below the DM contribution at $R\sim8\kpc$
and thus fulfils all microlensing constraints. Its Snhd $v_{\rm circ}(R_0)$ is lower
than that found by \citet{ralph2012}.

Model \WA1 has a thick IC disc with $M_{\rm disc,i}=2.0\times10^{10}\msun$
and $h_{R, {\rm disc}}=2.0\kpc$ and in addition an IC bulge with $a_{\rm bulge}=0.7\kpc$ and
$M_{\rm bulge,i}=0.5\times10^{10}\msun$. It has inside-out growth in the range $1.5-3.5\kpc$ 
and the same final mass as U1. Consequently, its peak $v_{\rm circ, b}$ is higher at
$\sim200\kms$, but it still fulfils all constraints from microlensing, as well
as the Snhd $v_{\rm circ}(R_0)$ constraints. The declining $\sigma_0$ model \V6 also fulfils
all constraints but in a different way. It starts from a low-mass elliptical IC
and grows a disc with $M_{\rm f}=6\times10^{10}\msun$ like those of two previous models.
Its inside-out growth is from $1.0$ to $3.5\kpc$ and its final $v_{\rm circ, b}(R)$
is rather constant at $170-180\kms$ in the range $R=2-10\kpc$. Unlike \WA1,
which has a flat total circular speed with $240\kms$ for $R=3-10\kpc$, \V6 has 
$v_{\rm circ}(R)$ increasing in this radial range from $220\kms$ to $250\kms$.

\section{Bar formation and evolution}
\label{sec:bar}

Paper 1 showed that models starting with a thin-disc IC and having no GMC heating
undergo strong bar activity from early times. GMC heating can delay and weaken bar formation and 
evolution and in extreme cases prevent the formation of a strong bar over cosmological time-scales.
In this section, we examine how this picture is modified by an old thick-disc component. 

\subsection{Bar strengths}

Figure \ref{barx} displays for 10 models the evolution of the $m=2$ Fourier amplitude $A_2$ (see Equation \ref{a2})
for all the stars within $R=3\kpc$. Model Y2 represents thin-disc-only models and we see
that $\ln(A_2)$ instantaneously increases to $-2.5$ as the addition of mass to the thin and 
compact IC disc makes the system develop non-axisymmetries. As discussed in Paper 2, the radial heating by GMCs 
at low disc mass and high SFR is important and delays the formation of a strong bar during the first
$3\gyr$ of evolution in Y2. This effect is enhanced in model Y1$\zeta$-, which has $\zeta=0.04$ and thus 
twice as many GMCs per unit mass of added stars: $\ln(A_2)$ is kept at $\sim -3$ until $t\sim7\gyr$, when it 
increases to $\sim -1.7$, indicating a rather weak bar.

The curve for model P2 is very different. P2 shares with Y2, the constant radial growth history $h_R(t)=2.5\kpc$, 
the final mass $M_{\rm f}=5\times10^{10}\msun$ and the shape of the SFH. Its IC disc is thicker, more extended
and more massive than the one in Y2 and the normalization of its SFH is thus lower.
The existence of a hot disc, which is stable against bar formation prevents the growth
of $A_2$, although stars are continuously added on cold orbits throughout the simulation.
Only at $t\sim7\gyr$ has enough thin disc been accumulated to make the composite system
unstable to bar formation.

Paper 1 showed that, in the absence of a thick disc, lowering the halo concentration from $c=9$ is 
problematic, as the system becomes more self-gravitating and bar unstable. U1 has a more massive and 
more compact IC thick disc than P2, lives in a $c=7.5$ halo and has an inside-out growth history. 
Despite the higher thick-disc central surface density and the lower halo concentration than P2, 
the thick disc still suppresses bar formation for $4\gyr$. 

\begin{figure*}
\vspace{-0.cm}
\hspace{-0.5cm}\includegraphics[width=18cm]{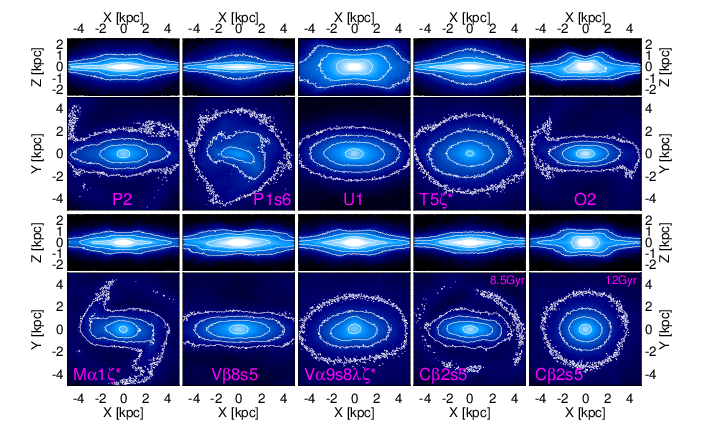}\\
\caption
{Edge-on and face-on surface density maps of the bar regions in several of our models. Most models
are shown at $t=t_{\rm f}$, only \C3b appears additionally at $t=8.5\gyr$.}
\label{barm}
\end{figure*}

Another ingredient here is the radial-to-vertical velocity dispersion ratio in the IC thick disc.
Model \PA1 has a more massive and more compact IC thick disc and a slightly higher
final mass than P2 and inside-out formation from $1.5$ to $3.5\kpc$. Despite the higher surface densities,
$\sigma_R^2/\sigma_z^2=1.8$ for the ICs in \PA1 compared to a value of 1 in P2 makes the system 
more stable against bar formation and thus weakens bar formation more strongly than in P2.

To understand bar evolution in models with declining $\sigma_0$, which
start with low-mass elliptical ICs, we first examine model K2 that grows a
thin disc inside a higher mass elliptical IC. K2 and P2 differ only in that
P2 has a thick-disc IC of the same mass. During the early evolution phases of
K2, $A_2$ is suppressed due to the elliptical ICs. However, in K2 $A_2$ increases
at an earlier time than in P2 and from $t\sim 3\gyr$ on shows a significant
bar. By construction, the in-plane velocity dispersions of the IC stars at $R\la 2\kpc$ 
are similar in these models. However, the surface densities are higher and thus the rotation
velocities of the IC stars are faster in K2. Consequently, K2 is more unstable 
to bar formation than P2.

Compared to the K ICs, the M ICs contain an elliptical, which is three times less massive.
So a cold disc model in M would have high $A_2$ at an earlier time than K2. The hot input dispersions for the old populations 
in M$\alpha$1, however, act in the same way as the thick IC disc in the P models and delay bar formation 
until $t\sim6\gyr$. We find that the specific shape of declining $\sigma_0(t)$ does not significantly 
influence bar formation.

V$\alpha$1 is the equivalent model to M$\alpha$1, but it lives in a lower concentration $c=6.5$ halo. Still,
bar formation is delayed until $t\sim4\gyr$. \V6 has a higher mass, a more compact feeding history and
a shorter SFR time-scale $t_{\rm SFR}$ than V$\alpha$1. All three factors lead to a much faster 
increase in surface density at early times, which outweighs the fact that at feeding the radial-to-vertical
input dispersion ratio $\sigma_R/\sigma_z=1.25$ is higher than in V$\alpha$1. Consequently, $A_2$ is higher 
at early times.

The specifics of the SFH, the radial growth history, the dispersions of the old components and GMC heating
thus determine the bar formation history of an individual model. Figures \ref{sd} and \ref{barx}
however show that halo concentrations $c=6-7$ allow models with reasonable final bars in the presence
of hot disc components. The fact that in the lower row of Figure \ref{sd}, the $m=2$ amplitudes
at final times and radii $R>5\kpc$ are lower in models with declining $\sigma_0$ than in the
thick IC disc models is connected to the gradients of $\sigma_R$. In models with thick-disc ICs,
the oldest stars have radially constant scaleheights and $\sigma_R/\sigma_z$ and thus declining $\sigma_R(R)$, 
whereas models with declining $\sigma_0$ assume a radially constant input dispersion $\sigma_R$.
Thus in the end, the outer thick components are radially hotter in models with declining $\sigma_0$
and thus less unstable to $m=2$ modes.

\subsection{Bar morphology}

The central region of our Galaxy is dominated by a bar, the inner part of which consists
of a boxy/peanut-shaped bulge at $R<2\kpc$  with an X-shape at $|z|>500\pc$ \citep{wegg13}
surrounded by a vertically thin part, the {\it long bar}, 
extending to $R\sim 4-5\kpc$ \citep{wegg15}. AS15 demonstrated that an inside-out growing model
without GMCs in a $c=9$ halo produces a bar with X-shaped structure with the tips of this structure
at $(x,z)\sim(2,1.3)\kpc$, very similar to the structure of the MW bulge/bar region inferred by 
\citet{wegg13}. Paper 1 showed that some thin-disc-only models with GMCs also displayed bars very
similar to the one in the MW, but also noted that not all of these bars are buckled
and that lower concentration haloes favoured unrealistically long bars.

Here we test how well our thick-disc models can reproduce the MW bar. This is interesting as the chemically defined
thick disc is concentrated and should thus have a high mass fraction in the bar region, but can only form
a bar if the thin-disc fraction is high enough, as discussed above. Figure \ref{barm} shows edge-on and
face-on surface density maps of several galaxies at $t_{\rm f}$.

Like the standard Y models in Paper 1 and the model in AS15, P2 lives in $c=9$ halo and shows a
bar that is $\sim 5\kpc$ long. Model P1s6, which has a more radially extended feeding history shows
only a small, weak bar in the central $R<2\kpc$. The vertical structure of P1s6 is indistinguishable
from a pure disc galaxy, whereas P2 has a boxy shape with a lateral extent of $\pm\sim 2\kpc$ and 
a vertical extent of $\pm\sim 1\kpc$ and a mild hint of an X structure. As in the MW bar, the outer
regions are thinner. \PA1 has a more compact and more massive, but radially hotter 
thick disc and does not show a bar, just mildly elliptical surface density contours.

The galaxy in model O2 evolves in a $c=9$ halo from an elliptical IC. At $t_{\rm f}$, its bar is $\sim 4\kpc$ long 
and is currently buckling as indicated by the broken mirror symmetry relative to the $x$-axis 
as first observed in a simulation by \citet{raha}. 
This event will eventually produce an edge-on peanut bulge with a characteristic X-shape (see \citealp{combes}).
U1 has a compact, massive IC thick disc, a higher-than-average final mass and lives in a $c=7.5$ halo. 
Its bar grows to a length $\sim 6\kpc$ and is thus only mildly longer than the P2 bar. However, bar buckling
has created a significantly more extended X-shape with a lateral extent of $\pm\sim 4\kpc$ and 
a vertical extent of $\pm\sim 2\kpc$.

The lower row displays bars in models with declining $\sigma_0$. Model \C3b grows a disc around a compact bulge IC
in a $c=9$ halo. At $t_{\rm f}$, it clearly shows an X-shaped edge-on structure in the central $\sim 2\kpc$,
stronger than the similarly sized one in P2. However, its face-on image reveals an almost axisymmetric image.
Going back in time, we find that at $t=8.5\gyr$ \C3b exhibited a strong bar that had not yet buckled.
From the evolution of $A_2(R<3\kpc)$ depicted in Figure \ref{barx}, we learn that the bar formed around $t=6\gyr$.
Between $t=8.5\gyr$ and $t_{\rm f}$ \C3b undergoes buckling, but in the final
$\sim 2\gyr$ of the simulation its bar becomes continuously weaker.

\begin{figure*}
\vspace{-0.5cm}
\hspace{-0.2cm}\includegraphics[width=14cm]{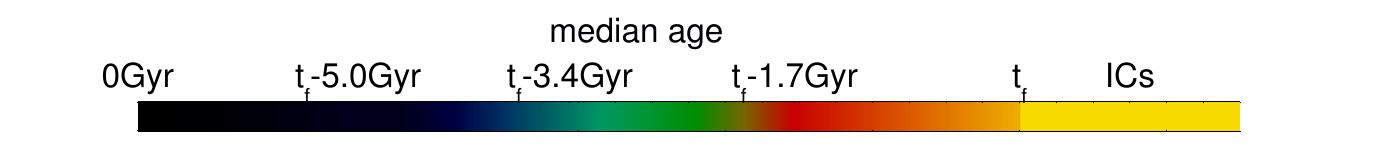}\vspace{-0.2cm}\\
\hspace{-0.5cm}\includegraphics[width=18cm]{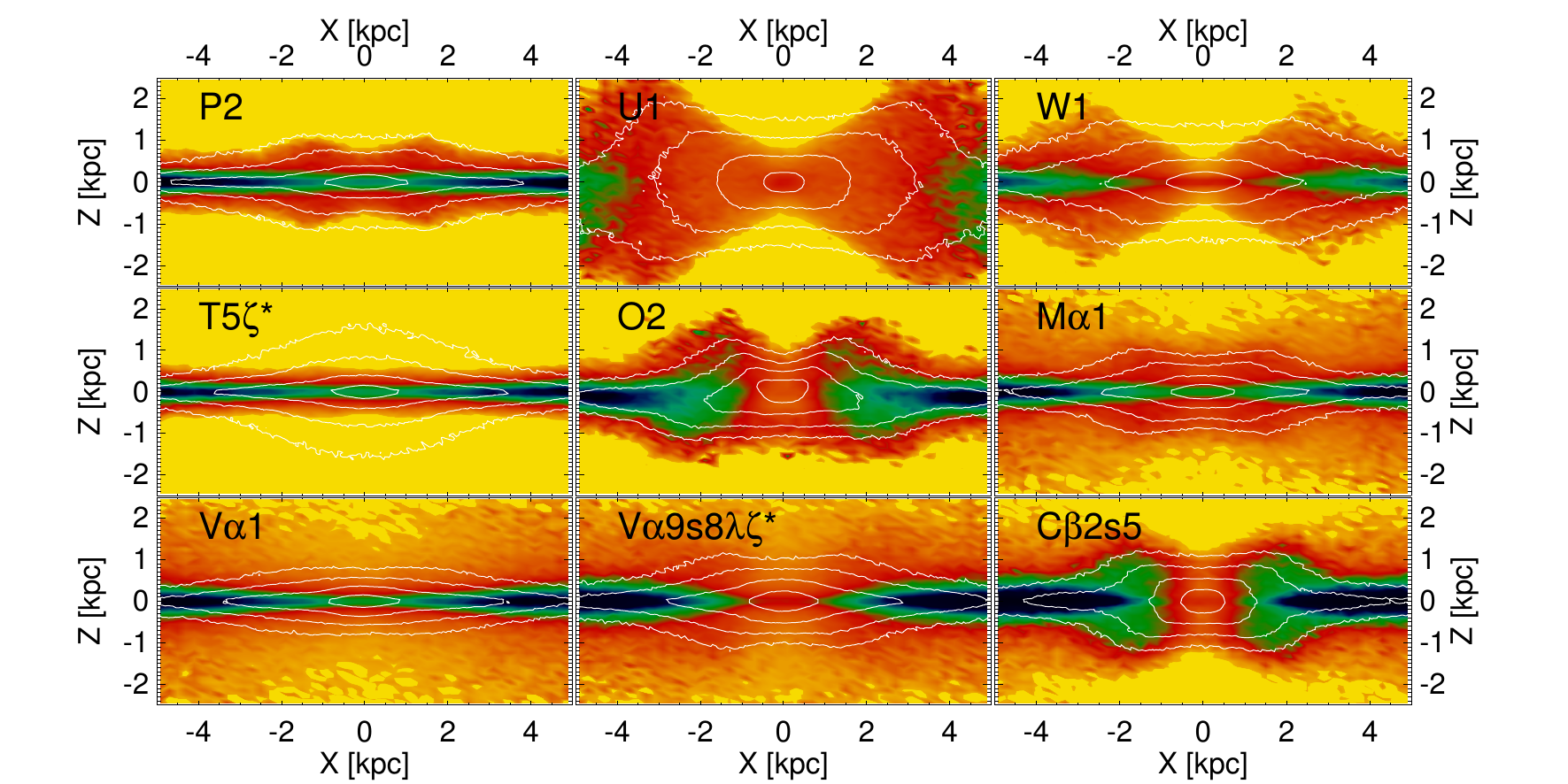}\\
\caption {Median ages $\tau_{\rm med}$ as a function of positions $(x,z)$ in edge-on
views of the central regions of several models. $x$ is along the major axis
of the bar.  Surface density contours are overplotted in white. Models are
shown at $t=t_{\rm f}$.} \label{bara}
\end{figure*}

Models M$\alpha$1$\zeta$* ($c=9$ halo) and V$\beta$8s5 and \V6 ($c=6.5$ halo) are models with declining
$\sigma_0$ starting from low-mass elliptical ICs. They display bars of reasonable sizes ($3.5-5.5\kpc$) and 
varying strengths. Their vertical profiles all show boxy edge-on shapes with vertical extents 
$\pm\sim1.0-1.5\kpc$ and lateral extents that are wider for longer bars.

In summary, it is possible to create models that show reasonable agreement with the vertical profile
of the MW, its circular speed curve and at the same time contain a bar, which in length, strength
and vertical extent agrees reasonably with that of the MW. The problem is that the details depend on
mass and size growth history, DM halo density and GMC heating and that, additionally, the evolution of 
bar length and strength is to some degree stochastic \citep{sellwood09}, so that, not even within the
limits of our methods, it is possible to determine which model best represents our Galaxy.

\subsection{Bar age structure}

In Figure \ref{bara}, we examine the age structure of the bar. For several models we plot a map
of median age $\tau_{\rm med}$ in the $x$-$z$ plane, where $x$ is measured along the major axis of the bar
and $z$ is perpendicular to the disc. All stars with $|y|<1.5\kpc$ are considered for the map.
If $\tau_{\rm med}$ at a certain position corresponds to a star particle from the ICs, we apply a distinct 
yellow colour, whereas inserted star ages range from black (young) to orange (old). We overplot
in white edge-on density contours.

We start by analysing model \PA1, which does not show a noticeable bar, in the left-hand panel 
of the middle row. Clearly, above $|z|\sim600\pc$ the IC thick disc dominates at all $x$. The thin
disc is youngest in the plane and has a vertical age gradient, as an effect of stars being born 
cold and being vertically heated by GMCs. It also has a radial age gradient, due to inside-out formation.

In contrast to model \PA1, the three models displayed in the the top row of Figure \ref{bara} at $t_{\rm f}$ 
show X-shaped edge-on structures. These three models all grow thin discs within thick IC discs. The X-shape is beautifully
visible in the $\tau_{\rm med}$ maps. The cones of relatively lower surface density above and below the galactic
centres are dominated by old, thick-disc stars, whereas younger stars fill diamonds that lie at both sides
of the cones in the $x$-direction. These diamond-shaped regions have vertical age gradients of different
strengths. This is connected to the fact that we do not feed new stars into regions of strong bars.
The bar in U1 forms already at $t=4\gyr$, whereas the bar in P2 only forms at $t=7.5\gyr$. As shown in AS15,
stars can be captured by the bar, but the rates of this process are low, so the age structure is only mildly
affected. Consequently, the number of relatively young stars present in P2 is high and vertical and radial
age gradients are present in the bar region. In U1, the region at $|x|<4\kpc$, which is part of the diamonds,
is well mixed in $\tau_{\rm med}$. The bar in W1 forms at $t=6\gyr$ and the model is thus intermediate to P2 and U1
in terms of age gradients.

The vertical mixing of stars of various ages happens during bar buckling, as is illustrated by model O2, which
is undergoing the process at the depicted moment. We see how the younger stars are spread vertically
at around $|x|\sim2\kpc$ and the model thus transitions from a disc-like to an X-shaped age structure.
Note that, as bars can vary in length and strength with time, cold, young stars can be added in the plane
after buckling events and there can be multiple buckling events, which complicates the age structure.

The remaining four models in Figure \ref{bara} feature declining $\sigma_0$. As the thick discs of these models
comprise fed-in stars, which have radial and vertical density distributions that vary continuously with age,
the age structure is more complicated to interpret, as in real galaxies. At $t_{\rm f}$, V$\alpha$1, like \PA1,
has no X-shaped edge-on structure. The thin discs of these galaxies exhibit qualitatively similar vertical and 
radial age gradients. The thick disc of V$\alpha$1 shows a continuation of these gradients at higher $|z|$; the
gradients in these regions are, however, much shallower. 

M$\alpha$1 and \V6 have edge-on peanut-shaped density structures similar to that of P2 and, indeed, the age structures of
the three bars are qualitatively similar. As the thick disc background, however, is, again, much less distinct, 
the structure combining old cones plus younger diamonds is much less evident and might be hard to measure in a real
galaxy.

Model \C3b differs from the other models with declining $\sigma_0$ in having compact bulge ICs and thus the density
of IC stars near the galactic centre is higher and the $\tau_{\rm med}$ gradients away from the centre are steeper
than in models with low-mass elliptical ICs. A relatively recent and strong buckling event before the 
disappearance of the bar and a radial growth history with constant $h_R=2.5\kpc$ has led to relatively young 
areas at $|x|\sim2\kpc$.

\section{Discussion}
\label{sec:discuss}

\subsection{Vertical profiles and age gradients}

In all final models the vertical profile of the Snhd is fitted well by the
sum of two exponentials similar to that of the MW. When the IC contains a
thick disc, the vertical profile is double-exponential from the outset, with a
growing thin-to-thick density ratio, whereas when $\sigma_0$ declines, the
vertical profile gradually evolves a double-exponential structure as a
sufficiently massive, cold and thin population forms, and the scaleheights of
both components change continuously. 

We cannot directly measure the evolution of scaleheights over cosmic time,
but in nearby galaxies we can probe this evolution through observations of
the radial variation of the vertical profile and age structure of the disc.
The radial variation of the vertical profile is dominated by the evolution of
the thick disc because our thin discs have scaleheights that are almost
independent of both $R$ and $t$. Our thick-disc ICs are set up with
declining $\sigma_z(R)$ and radially constant scaleheights, and the latter
property is roughly conserved to the present epoch, as radial migration causes
only mild levels of flaring. Models with declining $\sigma_0$ have, by 
construction, radially constant $\sigma_z$ at birth. As the oldest and
hottest stars are barely affected by vertical heating, their thick discs flare
strongly and their scaleheights increase with radius. In these models, radial
migration weakens flaring, but the effect is not strong enough to balance
the outward increase in scaleheight imprinted at birth.

This has important consequences on the age structure of the disc. At radii
$R=5-12\kpc$ and altitudes $|z|>1\kpc$ models with thick-disc ICs are
dominated by old IC disc stars; they thus show no radial variation in
median age $\tau_{\rm med}$. At lower altitudes these discs become younger with increasing
$R$.  The age structure is markedly different in models with declining
$\sigma_0$.  On account of the strongly flaring old and intermediate-age
components, they show negative $\tau_{\rm med}$ gradients at all $|z|$. Models with
declining $\sigma_0$ thus agree better with recent measurements of the radial
age structure at various altitudes in the MW by \citet{martig2}. They also
show somewhat flatter vertical age gradients at $R_0$, in rough agreement
with measurements by \citet{casagrande}. For thick IC disc models, strong
inside-out growth improves the agreement with these measurements.

Our discs evolve in isolation, whereas at least the low-density outskirts of
discs are likely to be affected by processes capable of significant vertical
thickening, such as disc-satellite interactions \citep{kazantzidis} or infall
of gas with misaligned angular momentum \citep{jiang}.  Including minor
mergers or adding stars in tilted outer discs would thus be a valuable
extension of our models.

\subsection{The interplay between the dark halo, the thick disc and the bar}

The evolution of a disc depends on the local density of the dark halo because
increasing the latter reduces the extent to which the disc controls the
gravitational field in which it moves. In particular, decreasing the local DM
density increases the amplitude of non-axisymmetric structure.  We varied the
local DM density by varying the initial concentration parameter $c$ at a
fixed halo mass, $M_{\rm tot}=10^{12}\msun$.  Concentrations in the range
$c=6-9$ work well. Indeed, after the DM has been compressed by the disc, the
final DM density in the Snhd then agrees with observational constraints, and
in many models a bar similar to that of the MW emerges before the current
epoch. Although non-axisymmetric structures in the disc transfer angular
momentum to the DM, the dark halo does not acquire a core like that favoured
in the MW \citep{cole}. If such a core exists in a $\Lambda$CDM context, it
thus probably formed in the very early evolution stages of the
Galaxy that are not modelled here. 

Observational constraints on the circular speed, $v_{\rm circ}(R_0)$, near
the Sun \citep{ralph2012} and on the microlensing optical depth towards the
MW bar/bulge \citep{wegg, cole} indicate a baryon-dominated central MW and
roughly equal contributions of DM and baryons to $v_{\rm circ}(R_0)$. To
achieve this in our models, initial concentrations $c\sim6-7$ are favoured as
higher values do not allow for enough baryonic mass in the central regions.

Paper 1 favoured $c\sim9$ because thin-disc-only models with lower values of
$c$ showed unrealistically long and strong bars.  The presence of old thick,
and thus kinematically hot, disc components alleviates this problem. By
shifting mass from the cold thin disc to a thick, radially hot component, the
formation of a bar can be delayed by several Gyr. It is immaterial whether
the hot component is included in the ICs or arises from declining $\sigma_0$,
and the ratio of radial to vertical dispersions can be
$\sigma_R/\sigma_z\sim1$, so smaller than the values $\sigma_R/\sigma_z\sim2$
in thin discs. As a consequence, the time available for the bar to grow in
strength and length is limited. Indeed, 
models with $c=9$ haloes tend to show an unrealistically weak final bar. Models with
$c\sim6-7$ generally have bars with lengths similar to that of the MW's bar.
The delaying of bar formation by old thick discs also explains the observation 
that the fraction of barred disc galaxies decreases with increasing redshift \citep{sheth}.

\subsection{The edge-on structure of bars}

Several models have bars that are morphologically similar to that of the MW:
a boxy bulge in the central $R\la2\kpc$ with an X-shape up to
$|z|\sim1-1.5\kpc$ at the centre of a thinner outer bar that extends to
$R\sim5\kpc$ \citep{wegg13,wegg15}. In one of these models the bar dissolved
at the very end of the simulation. However, notwithstanding the face-on
surface density being almost axisymmetric, the edge-on peanut shape survived.
Thus not all observed boxy edge-on bulges need be bars.

In edge-on density projections of these models, the X-shapes are not as
striking as in the model of AS15, which lacks a thick disc and GMCs, but
features an isothermal gas component. This is not surprising as the MW bulge
does not show an X-shape in all stellar components. \citet{dekany}
found that old and metal-poor RR Lyrae stars appear to have a more spheroidal
shape and \citet{portail17} found that low-metallicity stars ($\feh<-0.5$) 
in the bar/bulge contain a much lower fraction of stars on bar-supporting orbits 
than stars with higher metallicities. Moreover, the X is possibly absent in 
younger populations as well \citep{lopez}. Analysis on the edge-on age structure 
of our bars offers insight into why the shape should vary with stellar population.

The X structures are particularly pronounced in models with thick-disc ICs
that have final boxy bulge/bar regions that are more extended than in the
MW. Here bar buckling spreads stars from the thin disc vertically, and because
they mainly populate the 2:2:1 resonant orbit family \citep{pfenniger}, these 
stars form a structure that, seen edge-on, resembles two diamonds overlapping 
at the galactic centre. The cones above and below the centre, which are not 
populated by the 2:2:1 orbits, are dominated by thick-disc stars. Young stars 
that were captured by the bar after the buckling event will be found in the 
plane. Consequently, a distinct age pattern should be observed in buckled 
edge-on bars. 

 A similar separation in edge-on morphology between the oldest, 
the intermediate-age and the youngest stars in a barred galaxy has also been 
found in the simulations of \citet{atha}. \citet{debattista16} recently studied
bar formation in galaxies that contain disc populations with differing random 
motions. They showed that radially cooler populations form stronger bars, the edge-on 
profiles of which are vertically thinner and peanut-shaped, whereas the hotter 
populations form a weaker bar with a vertically thicker edge-on box shape (see also 
\citealp{fragkoudi}).

The thick and thin discs of models with declining $\sigma_0$ are not strictly
separated in age, as they are in models with thick-disc ICs, and in
consequence their characteristic age structure is less clear in an edge-on
age map. Thus while observations of the edge-on age structures of bars have
the potential to betray the formation history of the bulge, central thick and
thin discs and the timing of the bar buckling event, the constraints will be
less tight if the scenario with declining $\sigma_0$ is more appropriate than
that in which the thick disc is included in the ICs.

\subsection{Radial redistribution and inside-out growth}

Bars and spiral structure make it hard to steer the disc's radial scalelength 
$h_R$ at the solar radius $R_0$ to a preferred value, $h_R(R_0)\sim 2.6\kpc$. As was
already discussed in Paper 1, making the disc more compact results in stronger 
non-axisymmetries, which in turn leads to more mass redistribution and larger 
$h_R(R_0)$. We have shown that this is the case for all age components and for 
all models with appropriate bars, so in the past $h_R(R_0)$ would likely have 
been smaller than it is today. Assigning higher values of $\sigma_R/\sigma_z$ 
to the old thick disc can yield steeper-than-average profiles, but doing 
so weakens bars inappropriately. 

Still, for models that grow inside out, $h_R(R_0)$ always increases
with decreasing age, just as observations of the Snhd suggest \citep{bovy}.
As these observations show scalelengths $h_R(R_0)\sim1.5-2.0\kpc$ for the
most compact and oldest mono-abundance populations (see also
\citealp{cheng}), $h_R=2\kpc$ can be regarded as an upper limit on the birth
scalelength at the earliest times, but model \V6 demonstrates that the
scalelength at birth could have been as small as $h_R=1\kpc$, a conclusion
similar to that of \citet{ralph2017}.  The flat age-metallicity relation of
the Snhd and the radial metallicity gradient in the MW make it hard to infer
the scalelength of current star formation by studying mono-abundance
populations. In our models input scalelengths at late times in the range
$h_R\sim3-4 \kpc$ give reasonable results.

\subsection{Thick-disc formation scenarios}

The age structure of the MW disc points towards a model with declining
$\sigma_0$. Most hydrodynamical cosmological simulations of disc galaxies 
support a picture in which birth dispersions and gas fractions decline 
continuously with time (e.g.  \citealp{bird, stinson, ma}). Such a scenario
has also been inferred from observations of H$\alpha$ kinematics (e.g.\ 
\citealp{kassin, wisn}). However, \citet{diteodoro} argue that there is no 
substantial difference between the gas kinematics of galaxies at redshifts 
$z_{\rm rs}\sim1$ and today. Moreover, \citet{martig14} find that models that 
have kinematics in line with those found in the Snhd favour a two-phase 
formation scenario, in which the thick-disc stars are born in a turbulent, 
merger-dominated phase and the thin-disc stars are born cold and heated 
subsequently.

Such a two-phase scenario motivated our models with thick-disc ICs, but in
setting up an equilibrium thick stellar disc with a radially constant
scaleheight we have ignored correlations between stellar ages, kinematics and
density profiles that would naturally arise during formation of the
proto-thick disc.  Declining $\sigma_0$ inherently produces such correlations.
However, an inappropriate thick disc is still liable to emerge through poor
choices for $\sigma_0(t)$ or $h_R(t)$, or the choice of a radially constant
$\sigma_0$.

The main limiting factor of our models is thus the lack of a self-consistent
heating mechanism for thick-disc stars: in both scenarios thick-disc stars
are created ad hoc. The heating mechanism will affect non-axisymmetries and
the radial distribution of matter, which are crucial for disc evolution. As
measurements of gas fractions at redshifts $z_{\rm rs}\sim2$, a time
consistent with the formation of the chemically defined thick disc of the MW,
indicate that molecular gas makes up 50 per cent or more of the baryonic
masses of galaxies (e.g.\ \citealp{genzel15}), the lack of a realistic gas
component is a connected problem. Although at early times models such as \V6
have as much as 45 per cent of their baryonic mass in GMCs, the GMCs have the
same mass function as a present-day spiral galaxy, in which the gas
fraction is lower and stars form cold. In a picture in which turbulence
driven by gravitational disc instabilities causes stars to form with large
dispersions \citep{forbes}, molecular complexes would be expected to be more
massive.  Scattering of stars by massive clumps could contribute to thick-disc
formation \citep{bournaud}.

\section{ Conclusions} 
\label{sec:conclude}

We have presented a new set of idealized $N$-body simulations of disc
galaxies with both thin and thick discs within live dark haloes. These
models are grown over $10-12\gyr$ by continuously adding new stellar
particles with specified age-dependent velocity dispersions. Short-lived
massive particles represent GMCs. Thin discs grow by the addition of 
stars on near-circular orbits, whereas for thick-disc components we rely 
on two different concepts: a) create an appropriate thick disc in the 
ICs and only add thin-disc stars during the simulation, or b) start with 
low-mass, diffuse elliptical or compact bulge ICs and add stars with 
continuously declining input velocity dispersion $\sigma_0(t)$. Hence in 
scenario b) we form kinematically hot thick-disc stars at early times 
and cold thin populations at late times, whereas in scenario a) the 
simulation starts after the structure that will morph into the thick 
disc is fully formed. To understand the evolution of our models, we 
simulate a variety of histories of star formation, dark halo densities and
thick disc properties.

Both types of models can produce at final time $t_{\rm f}$ models that are
similar in structure to the MW. We find:
\begin{itemize}
\item{Both scenarios create double-exponential vertical profiles. The scaleheight
of the thin disc is governed by GMC heating. To achieve a MW-like thick-disc 
exponential scaleheight $h_{\rm thick}\sim 1\kpc$ at $t_{\rm f}$, thick-disc ICs with 
isothermal vertical scaleheights $z_0\sim1.7\kpc$ are suitable. For declining 
$\sigma_0$ models, the input velocity dispersions should be $\sigma_0\sim40-50\kms$
at the earliest formation stages. Thick-disc scaleheights are not affected by GMC
heating.}
\item{Models need to undergo inside-out growth to reproduce the observed dependence
of radial scalelength $h_R$ on chemical composition of disc stars. We find that models
that grow from $h_R\sim1-2\kpc$ at early times to $h_R\sim3-4\kpc$ today are suitable.}
\item{To explain the baryon dominance of the Galactic Centre, the circular speed curve
of the MW and the structure of the bar in the presence of a thick disc, we favour DM 
haloes that at mass $M_{\rm{DM}}=10^{12}\msun$ have an initial concentration parameter 
$c\sim7$.}
\end{itemize}

It is essential that thick-disc stars are already hot when the thin disc
starts forming because Paper 1 showed that heating by GMCs and
non-axisymmetries is incapable of producing the thick disc, although it explains the
properties of the thin disc. The presence of the thick disc modifies the
evolution of the thin disc, but the final properties of the thin discs
in our thin+thick disc models are similar to those of thin-disc-only models
in slightly more concentrated dark haloes. Crucially, this change in halo
density and the presence of a hot and thick disc make it possible to bring
models with appropriate bars into agreement with the baryon fractions
inferred for the central MW.

Regarding the non-axisymmetric structures of the disc, we find:
\begin{itemize}
\item{Bars with a structure similar to that of the MW bar, i.e. a
boxy/peanut-shaped bulge at $R<2\kpc$ with an X-shape surrounded by a
vertically thin part extending to $R\sim 4-5\kpc$, can be found in some of
our viable models. Stochasticity in the evolution of bar lengths and
strengths complicates the comparison.}
 \item{The presence of a hot, thick-disc stellar population at the start of
thin-disc formation suppresses non-axisymmetries and delays the formation of
the bar.}
 \item{In models with an appropriate bar, the local exponential scalelengths
$h_R(R_0)$ of all mono-age populations are increased by the radial
redistribution of matter that the bar and spirals generate. Populations
measured in the Snhd today have thus likely had lower $h_R(R_0)$ in the
past.}
 \item{The dark halo's density profile is modified by the growth of the disc
and the non-axisymmetric structures that form in the disc, but the profile
does not develop a central core as is currently favoured for the centre of
the MW.}
\end{itemize}

To distinguish between formation scenarios, it is helpful to study the radial and 
vertical age structure of disc galaxies. We find:
\begin{itemize}
\item{Our two types of models for the creation of the thick disc differ
significantly in the predicted age maps of the discs. The observed Snhd
radial age gradient at $|z|>1\kpc$ and the vertical age gradient both favour
models with declining $\sigma_0$. However, the measurements are still rather
uncertain and models with thick IC discs by construction ignore any internal
structure of the thick disc. Such structure could be added to these models.}
 \item{Bar buckling in thin+thick disc systems creates characteristic age
patterns in edge-on views of the bar region. Buckling predominantly affects
thin-disc stars and causes them to form a structure in the $(R,z)$ plane
resembling two diamonds overlapping at the galactic centre. The cones above
and below the centre are dominated by thick-disc stars.}
\end{itemize}

Considering the wealth of data on the structure of the MW that will soon become available
from surveys such as Gaia \citep{gaia}, evolutionary models of disc galaxies that grow over
cosmological time-scales and contain both thick and thin discs will be essential to
connect the data to the formation history of the MW. Our models allow for a relatively
controlled and flexible setup, can be produced in large numbers and capture a wealth of
important dynamical processes.

We have demonstrated that our models can reasonably reproduce a variety of observations
of the structure of the MW. No model sticks out as particularly similar to the MW in all 
aspects, but this is  to be expected given the remaining shortcomings in 
modelling. In a companion paper (Paper 4) we examine the models presented here in 
light of Snhd kinematics and constraints on radial migration.

\section*{Acknowledgements}
We thank the referee for comments that helped improve the paper.
It is a pleasure to thank Ralph Sch{\"o}nrich for valuable discussions
and comments on the manuscript.

This work was supported by the UK Science and Technology Facilities Council (STFC)
through grant ST/K00106X/1 and by the European Research Council under the European 
Union's Seventh Framework Programme (FP7/2007-2013)/ERC grant agreement no.~321067.
This work used the following compute clusters of the STFC DiRAC HPC Facility 
(www.dirac.ac.uk): i) The COSMA Data Centric system at Durham University, operated
by the Institute for Computational Cosmology. This equipment was funded by a BIS 
National E-infrastructure capital grant ST/K00042X/1, STFC capital grant 
ST/K00087X/1, DiRAC Operations grant ST/K003267/1 and Durham University. 
ii) The DiRAC Complexity system, operated by the University of Leicester 
IT Services. This equipment is funded by BIS National E-Infrastructure capital 
grant ST/K000373/1 and STFC DiRAC Operations grant ST/K0003259/1.
iii) The Oxford University Berg Cluster jointly funded by STFC, the Large 
Facilities Capital Fund of BIS and the University of Oxford. 
DiRAC is part of the National E-Infrastructure.

\label{lastpage}
\end{document}